\documentclass[twocolumn,10pt]{IEEEtran}
\usepackage{stfloats}
\usepackage{xcolor}
\usepackage{amssymb}
\usepackage{theorem,amsbsy,pifont,verbatim,amsfonts,amssymb,color}
\usepackage{xspace,epsfig,graphicx,subfigure,latexsym,fancyhdr,cite}
\usepackage{multirow}
\usepackage{epstopdf}
\usepackage{makecell}
\usepackage{bm}
\usepackage{booktabs}
\usepackage{indentfirst}
\usepackage{cuted}
\usepackage{amsmath}
\usepackage{diagbox}
\usepackage{enumerate}
\usepackage{float}
\usepackage{url}
\usepackage{mdwmath}
\usepackage{mdwtab}
\usepackage{amsmath}
\usepackage{xcolor}
\usepackage{caption}
\usepackage{graphfig}
\usepackage{enumitem}
\usepackage{array}
\usepackage{tabularx}
\usepackage[colorlinks,citecolor=blue,linkcolor=blue,urlcolor=blue]{hyperref}
\usepackage{lipsum}

\def\0{\boldsymbol 0}
\def\1{\boldsymbol 1}

\allowdisplaybreaks

\ifCLASSINFOpdf
\else
\fi

\usepackage{stfloats}

\begin{document}
	%
	\title{Beyond $\lambda/2$: Can Arbitrary EMVS Arrays Achieve Unambiguous NLOS Localization?}
	%
	%
	%
	%

	\author{Hua~Chen,~
		Zhenhao Yu,
		Tuo Wu,
		Wei~Liu,
		Maged Elkashlan,
		Hyundong Shin, \emph{Fellow,~IEEE},\\
	    Matthew  C. Valenti,~\IEEEmembership{Fellow,~IEEE},  
		and~Robert Schober,~\IEEEmembership{Fellow,~IEEE}
		
		\thanks{This work was supported by by the National Natural Science Foundation of China under Grant 62571286, by the ``Pioneer" and ``Leading Goose" R \& D Program of Zhejiang Province under Grant 2024C01105, the Zhejiang Provincial Natural Science Foundation of China under Grant LY23F010003, and by the China Scholarship Council under Grant 202408330215.  (\emph{Corresponding author: Tuo Wu}.)}
		\thanks{Hua Chen, Zhenhao Yu are with the Faculty of Electrical Engineering and Computer Science, Ningbo University, Ningbo 315211, China. (E-mail: $\rm dkchenhua0714@hotmail.com; 15757087387@163.com$). T. Wu  is with the School of Electronic and Information Engineering, South China University of Technology, Guangzhou 510640, China (E-mail: $\rm wtopp0415@163.com$).Wei Liu is with the Department of Electrical and Electronic Engineering, The Hong Kong Polytechnic University, Hong Kong. (E-mail: $\rm  wei2.liu@polyu.edu.hk$). M. Elkashlan is with the School of Electronic Engineering and Computer Science at Queen Mary University of London, London E1 4NS, U.K. (E-mail: $\rm maged.elkashlan@qmul.ac.uk$). H. Shin is with the Department of Electronic Engineering, Kyung Hee University, Yongin-si, Gyeonggi-do 17104, Korea (E-mail: $\rm hshin@khu.ac.kr$). M. C. Valenti is with the Lane Department of Computer Science and Electrical Engineering, West Virginia University, Morgantown, USA (E-mail: $\rm valenti@ieee.org$). R. Schober is with the Institute for Digital Communications, Friedrich Alexander-University Erlangen-N$\ddot{\textrm u}$rnberg (FAU), Germany.  (E-mail: $\rm robert.schober@fau.de$).}
	}

	\markboth{IEEE Transactions on Signal Processing~Vol.~XX, No.~XX, XX~2026}
{Shell \MakeLowercase{\textit{et al.}}: A Sample Article Using IEEEtran.cls for IEEE Journals}
\maketitle

	\maketitle
	
	\begin{abstract}
Conventional radar array design mandates interelement spacing not exceeding half a wavelength ($\lambda/2$) to avoid spatial ambiguity, fundamentally limiting array aperture and angular resolution. This paper addresses the fundamental question: \textit{Can arbitrary electromagnetic vector sensor (EMVS) arrays achieve unambiguous reconfigurable intelligent surface (RIS)-aided localization when element spacing exceeds $\lambda/2$?} We provide an affirmative answer by exploiting the multi-component structure of EMVS measurements and developing a synergistic estimation and optimization framework for non-line-of-sight (NLOS) bistatic  multiple input multiple output (MIMO) radar. A third-order parallel factor (PARAFAC) model is constructed from EMVS observations, enabling natural separation of spatial, polarimetric, and propagation effects via the trilinear alternating least squares (TALS) algorithm. A novel phase-disambiguation procedure leverages rotational invariance across the six electromagnetic components of EMVSs to resolve $2\pi$ phase wrapping in arbitrary array geometries, allowing unambiguous joint estimation of two-dimensional (2-D) direction of departure (DOD), two-dimensional direction of arrival (DOA), and polarization parameters with automatic pairing. To support localization in NLOS environments and enhance estimation robustness, a reconfigurable intelligent surface (RIS) is incorporated and its phase shifts are optimized via semidefinite programming (SDP) relaxation to maximize received signal power, improving signal-to-noise ratio (SNR) and further suppressing spatial ambiguities through iterative refinement. Theoretical analysis establishes the maximum number of identifiable targets ($K = N$ for $N$ receivers) and derives the Cram\'er-Rao bound (CRB) as a performance benchmark. Simulation results demonstrate near-CRB accuracy even when interelement spacing significantly exceeds $\lambda/2$, confirming that arbitrary EMVS arrays can indeed achieve unambiguous localization beyond the traditional $\lambda/2$ constraint.
	\end{abstract}
	
	\begin{IEEEkeywords}
		EMVS-MIMO, DOA, Polarization, RIS, SDP.
	\end{IEEEkeywords}

	%
	\IEEEpeerreviewmaketitle

	\section{Introduction}
	%
	%
	%
	
	
	Multiple-input multiple-output (MIMO) radar systems have garnered considerable attention in the field of array signal processing due to the waveform and spatial diversity techniques they enable. These systems exhibit multifaceted advantages, including high target identifiability, enhanced angular resolution, and superior anti-jamming capabilities \cite{1,2,3,4}. MIMO radar can be broadly categorized into monostatic and bistatic configurations \cite{5}, with bistatic MIMO radar offering unique advantages in specific application scenarios such as forward-scatter detection and distributed sensing. As one of the core technologies in MIMO radar, direction of arrival (DOA) estimation has recently attracted significant research interest, particularly regarding high-resolution algorithms and optimization frameworks. However, a fundamental constraint has long governed array design in radar systems: to avoid spatial ambiguity, conventional wisdom dictates that inter-element spacing must not exceed half a wavelength ($\lambda/2$). This constraint, while ensuring unambiguous angle estimation, severely limits the achievable array aperture and consequently restricts angular resolution, especially when the number of physical sensors is limited by cost or platform constraints.
	
	\textit{Why go beyond $\lambda/2$?} The answer lies in the fundamental trade-off between array aperture and spatial resolution. For a fixed number of sensors $N$, restricting spacing to $\lambda/2$ limits the maximum aperture to $(N-1)\lambda/2$, resulting in angular resolution proportional to $1/N$. In contrast, arbitrary array geometries with inter-element spacing exceeding $\lambda/2$ can achieve apertures several times larger, offering proportionally finer angular resolution—critical for distinguishing closely-spaced targets or enhancing localization accuracy in radar applications. Moreover, arbitrary geometries provide deployment flexibility essential for platforms with irregular surfaces or space constraints, such as unmanned aerial vehicles, conformal arrays on aircraft fuselages, or distributed sensor networks. The central question thus emerges: \textit{Can we break free from the $\lambda/2$ shackles while maintaining unambiguous parameter estimation?} Since the introduction of subspace methods such as MUSIC \cite{6} and ESPRIT \cite{7,8}, significant progress has been made in DOA estimation for MIMO radar. Dimensionality reduction techniques \cite{9,10} and nuclear norm minimization \cite{11} have further improved computational efficiency. However, these methods predominantly assume uniform arrays with $\lambda/2$ spacing, leaving the ambiguity resolution challenge for arbitrary geometries largely unaddressed.
	
	Electromagnetic vector sensors (EMVS) offer a promising pathway toward answering this question. Unlike scalar sensor arrays (SCSA) that measure only signal amplitude \cite{6,7,8,9,10,11}, EMVS comprise six orthogonally oriented electromagnetic components co-located at a single spatial point, simultaneously capturing all electric and magnetic field components \cite{12,13}. This multi-dimensional measurement provides significantly more degrees of freedom, enabling EMVS arrays to achieve superior parameter estimation with fewer physical elements.{ EMVS elements are more complex than scalar sensors and may increase per-element hardware and calibration cost. However, EMVS provides multiple measurements per spatial location, which can reduce the number of required physical elements and relax $\lambda/2$ spacing constraints, potentially lowering overall array size and deployment cost for a given localization performance.}Moreover, the polarization diversity inherent in EMVS measurements offers an additional dimension for resolving ambiguities that arise in sparse arrays: targets sharing similar spatial signatures can be distinguished through their distinct polarization states. Early EMVS research focused on uniform geometries with $\lambda/2$ spacing \cite{14,15,16,17,18}, where the vector cross-product (VCP) technique and ESPRIT-based methods \cite{18} were shown to provide unambiguous solutions. Recent advances include compressive sensing \cite{19}, PARAFAC decomposition \cite{20}, and propagator methods \cite{21}. For arbitrary geometries, the authors of \cite{22} employed VCP with least squares, while \cite{23} proposed spatial smoothing for coherent targets. However, a critical limitation persists: existing arbitrary-geometry methods either sacrifice accuracy (e.g., VCP-based coarse estimation) or assume specific structural constraints, failing to provide a general solution for unambiguous localization beyond $\lambda/2$ spacing.
	
	The challenge intensifies when transitioning from line-of-sight (LOS) to non-line-of-sight (NLOS) scenarios. All aforementioned methods assume direct propagation paths between transmitters, targets, and receivers. When severe obstacles block these paths, conventional algorithms become inapplicable. In recent years, reconfigurable intelligent surfaces (RISs) have emerged as a transformative technology for NLOS radar \cite{24,25,26,27,34}. An RIS consists of a large number of passive reflecting elements that independently impose phase shifts on incident signals, establishing virtual propagation paths between NLOS targets and transceivers. The critical insight is that RIS integration with EMVS-MIMO radar creates a unique opportunity to address the "beyond $\lambda/2$" question in NLOS environments. Specifically, RIS offers three synergistic advantages: \textit{First}, RIS enables NLOS operation by creating controllable reflection channels \cite{25,26,27}, extending radar capabilities to obstructed scenarios where large-aperture arrays are most valuable for compensating reduced SNR. \textit{Second}, smart RIS phase design provides an additional optimization dimension: by coherently combining reflected signals \cite{28,29}, the system can enhance received SNR and shape the effective channel to favor specific array manifold structures, potentially mitigating spatial ambiguities. \textit{Third}, fully-polarized RISs equipped with EMVS elements \cite{31} enable joint spatial-polarimetric optimization, exploiting all six electromagnetic components to create disambiguation cues unavailable in scalar systems. Preliminary work has explored RIS-aided MIMO radar \cite{30,31,35,36,37,38,39}, but the fundamental question was remained unanswered: \textbf{Can arbitrary EMVS arrays achieve unambiguous RIS-aided localization when element spacing exceeds $\lambda/2$?}
	
	Answering this question affirmatively requires overcoming four fundamental challenges. \textit{First}, spatial ambiguity resolution: when element spacing exceeds $\lambda/2$, phase measurements wrap modulo $2\pi$, creating multiple feasible angle hypotheses. For arbitrary geometries lacking regular structure, classical disambiguation methods (e.g., Chinese Remainder Theorem for uniform spacing) fail. A novel mechanism exploiting EMVS's multi-component measurements is needed. \textit{Second}, RIS-induced model complexity: RIS reflection introduces additional phase modulation coupled with unknown target parameters, creating a bilinear observation model. Existing EMVS estimation algorithms \cite{18,19,20,21,22,23} assume passive propagation and suffer model mismatch in RIS-aided scenarios. Decoupling RIS effects from target signatures while preserving estimation accuracy poses a theoretical challenge. \textit{Third}, non-convex joint optimization: RIS phase shifts and target parameters are mutually dependent—optimal phases require knowledge of target locations, yet accurate localization benefits from optimized phases. This coupling yields a non-convex optimization landscape with unit-modulus constraints on RIS elements \cite{30,31}, computationally intractable for arbitrary arrays. \textit{Fourth}, polarimetric exploitation: while fully-polarized RISs \cite{31} provide six-dimensional electromagnetic control, systematically leveraging this capability for disambiguation in arbitrary geometries with NLOS propagation remains unexplored.
	
	This paper provides an affirmative answer to the titular question by proposing a comprehensive framework that achieves unambiguous localization for arbitrary EMVS arrays beyond $\lambda/2$ spacing in RIS-aided NLOS scenarios. The key innovation lies in a synergistic integration of tensor decomposition and convex optimization, specifically tailored to exploit EMVS's multi-dimensional structure. Our solution addresses the four challenges through two mutually reinforcing mechanisms: \textit{(i) PARAFAC-based estimation with phase disambiguation}: We construct a third-order tensor from EMVS measurements that naturally decouples RIS phase shift effects from target spatial-polarimetric signatures via parallel factor (PARAFAC) decomposition. A novel disambiguation procedure then exploits rotational invariance across EMVS's six electromagnetic components to resolve $2\pi$ phase ambiguities, enabling unambiguous direction-of-arrival/direction-of-departure(DOA/DOD) recovery even when spacing exceeds $\lambda/2$ in arbitrary configurations. \textit{(ii) SDP-relaxed RIS phase optimization}: We formulate RIS phase shift design as a signal power maximization problem and develop a semidefinite programming (SDP) relaxation that transforms the non-convex unit-modulus constrained optimization into a tractable convex program. By leveraging estimated parameters to guide phase shift design, the optimized RIS configuration enhances SNR and further suppresses ambiguities, establishing an iterative refinement loop that progressively improves localization accuracy.
	The main contributions with respect to the titular question are:
	\begin{itemize}
		\item[1)] \textit{Affirmative answer to the "beyond $\lambda/2$" question}: We prove, both theoretically and experimentally, that arbitrary EMVS arrays can achieve unambiguous RIS-aided localization with inter-element spacing exceeding $\lambda/2$. A unified signal model explicitly incorporates RIS phase shift modulation, arbitrary geometries, and full electromagnetic polarization. The third-order PARAFAC tensor structure naturally decouples RIS effects from target parameters, providing a principled foundation that resolves the model mismatch challenge.{An additional benefit of the proposed framework is improved robustness to carrier frequency variation. Since unambiguous localization is achieved without enforcing $\lambda/2$ inter-element spacing, a fixed EMVS array can operate reliably across a wider range of wavelengths than conventional array designs.}
		\item[2)] \textit{Phase disambiguation via EMVS rotational invariance}: We develop a novel disambiguation procedure that exploits rotational invariance relationships across EMVS's six electromagnetic components to resolve $2\pi$ phase wrapping in arbitrary geometries. This mechanism enables unambiguous joint recovery of 2-D DOD, 2-D DOA, and polarization parameters with automatic pairing, even when spacing significantly exceeds $\lambda/2$. Closed-form solutions via trilinear alternating least squares(TALS), ESPRIT, and VCP ensure computational efficiency.
		\item[3)] \textit{Convex RIS phase shift optimization for ambiguity suppression}: We formulate the RIS phase shift design as a non-convex unit-modulus constrained problem and develop an SDP relaxation that yields a tractable convex program. The optimized phase shifts maximize the received signal power based on the estimated parameters, simultaneously enhancing SNR and shaping the channel to suppress spatial ambiguities. Theoretical analysis establishes rank-one solution conditions and characterizes the performance gain enabled by RIS optimization.
	\end{itemize}
	\section{Preliminaries}
    \label{sec:system_model}
	This section establishes the mathematical foundations for achieving unambiguous localization beyond the traditional $\lambda/2$ spacing constraint. We first introduce the EMVS structure, whose six-dimensional electromagnetic measurements provide the multi-component information essential for phase disambiguation. We then formulate the RIS-aided signal model for arbitrary array geometries, explicitly incorporating the phase relationships that enable resolution of $2\pi$ ambiguities.
	
	\subsection{Structure of EMVS}
	
	The key to breaking the $\lambda/2$ barrier lies in exploiting the rich structure of EMVS measurements. An EMVS is a polarization-sensitive sensor comprising three orthogonally oriented electric dipoles and three orthogonally oriented magnetic loops that are co-located. This configuration enables independent measurement of electric and magnetic field components, providing six complex-valued observations per sensor—significantly expanding the information available for disambiguation compared to scalar sensors that yield only a single measurement. Assuming a transverse electromagnetic (TEM) plane wave impinges on the EMVS, the polarization-related spatial response vector can be expressed as:
	
	\begin{equation}
		\mathbf{b} \triangleq
		\begin{bmatrix}
			\cos(\theta)\cos(\varphi)\sin(\zeta)e^{j\varrho} - \sin(\theta)\cos(\zeta) \\
			\sin(\theta)\cos(\varphi)\sin(\zeta)e^{j\varrho} + \cos(\theta)\cos(\zeta) \\
			-\sin(\varphi)\sin(\zeta)e^{j\varrho} \\
			-\sin(\theta)\sin(\zeta)e^{j\varrho} - \cos(\theta)\cos(\varphi)\cos(\zeta) \\
			\cos(\theta)\sin(\zeta)e^{j\varrho} - \sin(\theta)\cos(\varphi)\cos(\zeta) \\
			\sin(\varphi)\cos(\zeta)
		\end{bmatrix}
		\label{eq:1}
	\end{equation}
	where \(\theta \in [-\pi, \pi]\), \(\varphi \in [0, \pi]\), \(\zeta \in [0, \pi/2]\), and \(\varrho \in [-\pi, \pi]\) denote the azimuth angle, elevation angle, polarization auxiliary angle, and polarization phase difference, respectively. It should be noted that the vector in \eqref{eq:1} is valid only for an ideal omnidirectional EMVS. Furthermore, it can be decomposed into a direction-dependent matrix \(\mathbf{V} \in \mathbb{C}^{6 \times 2}\) and a polarization-dependent vector \(\mathbf{g} \in \mathbb{C}^{2 \times 1}\) as follows:
	
	\begin{equation}
		\mathbf{b} = \mathbf{Vg}
	\end{equation}
	
	\noindent where
	
	\begin{equation}
		\mathbf{V} =
		\begin{bmatrix}
			\cos(\theta)\cos(\varphi) & -\sin(\theta) \\
			\sin(\theta)\cos(\varphi) & \cos(\theta) \\
			-\sin(\varphi) & 0 \\
			-\sin(\theta) & -\cos(\theta)\cos(\varphi) \\
			\cos(\theta) & -\sin(\theta)\cos(\varphi) \\
			0 & \sin(\varphi)
		\end{bmatrix},
		\mathbf{g} =
		\begin{bmatrix}
			\sin(\zeta)e^{j\varrho} \\
			\cos(\zeta)
		\end{bmatrix}
		\label{eq:3}
	\end{equation}
	
	Furthermore, vector $\mathbf{b} \triangleq \begin{bmatrix} b(1), b(2), b(3),b(4), b(5), b(6) \end{bmatrix}^T$ can be partitioned into two components, where the first three elements correspond to the electric-field vector $\mathbf{e} \in \mathbb{C}^{3 \times 1}$, while the last three elements correspond to the magnetic-field vector $\mathbf{m} \in \mathbb{C}^{3 \times 1}$, given by
	\begin{equation}
		\mathbf{e} \triangleq \begin{bmatrix} b(1), b(2), b(3) \end{bmatrix}^T
	\end{equation}
	\begin{equation}
		\mathbf{m} \triangleq \begin{bmatrix} b(4), b(5), b(6) \end{bmatrix}^T
	\end{equation}
	
	Moreover, vectors $\mathbf{e}$ and $\mathbf{m}$ satisfy the following fundamental electromagnetic constraint:
	\begin{equation}
		\mathbf{q} \triangleq \begin{bmatrix} \cos(\theta)\sin(\varphi) \\ \sin(\theta)\sin(\varphi) \\ \cos(\varphi) \end{bmatrix} \triangleq \frac{\mathbf{e}}{\|\mathbf{e}\|_F^2} \circledast \frac{\mathbf{m}^*}{\|\mathbf{m}\|_F^2}
		\label{eq:EMVS6}
	\end{equation}
	
	Evidently, the azimuth and elevation angles can be uniquely determined from \eqref{eq:EMVS6}. This vector cross-product relationship forms the basis of the VCP method, which provides coarse direction estimates without requiring phase information across array elements. However, as we will demonstrate, fully exploiting the rotational invariance relationships between $\mathbf{e}$ and $\mathbf{m}$ across multiple EMVS elements enables phase disambiguation for arrays with arbitrary spacing beyond $\lambda/2$.
	
	\subsection{RIS-aided Bistatic EMVS-MIMO Radar}
	\begin{figure}[!t]
		\centering
		\centerline{\includegraphics[width=9cm]{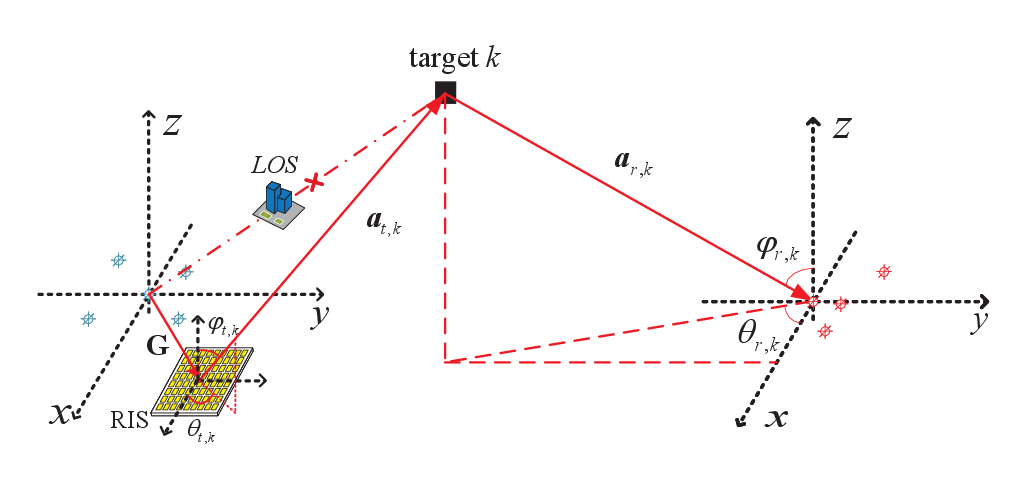}}
		\caption{The RIS-aided EMVS-MIMO radar model.}
		\label{fig:1}
	\end{figure}
	As illustrated in Fig. \ref{fig:1}, consider a bistatic MIMO radar model, where the transmitter and receiver employ arbitrarily configured arrays having $M$ and $N$ EMVS elements, respectively. The term ``arbitrarily configured'' is deliberate: unlike conventional systems constrained to $\lambda/2$ spacing for unambiguous estimation, our framework accommodates sensor placements with inter-element distances exceeding $\lambda/2$, thereby enabling large-aperture arrays without densely packing physical elements. Assume that $K$ non-coherent narrowband far-field targets exist in space, where the LOS path between the targets and the transmitter is obstructed, relying solely on the RIS to provide an NLOS path.{Here, non-coherence means that the signals backscattered from different targets are mutually uncorrelated, i.e., their initial phases are random and independent.} The RIS is configured as a nonuniform planar array with $Q$ elements, whose positions are $\mathbf{p}_q^s = [x_q, y_q, z_q]$. The position coordinates of transmit EMVS elements, targets, and receive EMVS elements are given by $\mathbf{p}_m^t = [x_m, y_m, z_m]$, $\mathbf{p}_k = [x_k, y_k, z_k]$, and $\mathbf{p}_n^r = [x_n, y_n, z_n]$, respectively. Critically,the $\mathbf{p}_n^r$ are not restricted to satisfy $\|\mathbf{p}_n^r - \mathbf{p}_{n'}^r\| \leq \lambda/2$ for all $n, n'$, allowing flexible deployment while phase disambiguation resolves resulting ambiguities. The azimuth and elevation angles are denoted as $\theta_{t,k}$ and $\varphi_{t,k}$ (DOD) for the $k$-th target relative to the RIS reference element, and $\theta_{r,k}$, $\varphi_{r,k}$ (DOA) for the receive array reference element. {Each transmit element emits orthogonal pulsed waveforms with equal unit power, where  the narrowband normalized  incident electric field signal of the TEM wave transmitted by the m-th EMVS, reflected via RIS towards the k-th target, is expressed as:}
	\begin{equation}
		\begin{aligned}
			\mathbf{r}_{m,k} &= \mathbf{V}_{t,k}^T \mathbf{d}_m \left[ \left[ \mathbf{G} \right]_m \text{diag}\{\mathbf{w}\} \mathbf{a}_{t,k} \right]^T \mathbf{s}_m(t) \\
			&= \boldsymbol{\zeta}_k \left[ \left[ \mathbf{G} \right]_m \text{diag}\{\mathbf{w}\} \mathbf{a}_{t,k} \right]^T \mathbf{s}_m(t) \\
			&= \begin{bmatrix} \boldsymbol{\zeta}_{k,H} \\ \boldsymbol{\zeta}_{k,V} \end{bmatrix} \left[ \left[ \mathbf{G} \right]_m \text{diag}\{\mathbf{w}\} \mathbf{a}_{t,k} \right]^T \mathbf{s}_m(t)
		\end{aligned}
	\end{equation}
	where $t$ denotes the time index, $\mathbf{V}_{t,k} \in \mathbb{C}^{6 \times 2}$ represents the directional matrix of the $k$-th target relative to the RIS, composed of the azimuth angle $\theta_{t,k}$ and elevation angle $\varphi_{t,k}$ as defined in \eqref{eq:3}. Additionally, $\mathbf{d}_m \in \mathbb{C}^{6 \times 1}$ denotes the polarization weight vector controlling the transmitting waveform. $s_m(t)$ denotes the waveform transmitted by the $m$-th EMVS, $\boldsymbol{\zeta}_k$ represents the probing polarization of the transmit EMVS array, while $\boldsymbol{\zeta}_{k,H}$ and $\boldsymbol{\zeta}_{k,V}$ denote the horizontal and vertical components of the waveform, respectively. $\left[ \mathbf{G} \right]_m$ represents the channel matrix from the $m$-th transmit element to the RIS, defined as:
	\begin{equation}
		\left[ \mathbf{G} \right]_m = \begin{bmatrix} e^{-j2\pi \rho_{m,1}/\lambda} & e^{-j2\pi \rho_{m,2}/\lambda} & \cdots & e^{-j2\pi \rho_{m,Q}/\lambda} \end{bmatrix}
	\end{equation}
	where $\lambda$ is the wavelength, $\rho_{m,q} = |\mathbf{p}_q^s - \mathbf{p}_m^t|$ denotes the distance between the $m$-th transmitting element and the $q$-th RIS reflecting element. $\mathbf{w}$ represents a RIS-related vector reflecting the amplitude variation$ \alpha_q$ and phase shift $\phi_q$ of the RIS elements, defined as:
	\begin{equation}
		\mathbf{w} = \begin{bmatrix} \alpha_1 e^{j\phi_1}, \ldots, \alpha_Q e^{j\phi_Q} \end{bmatrix}^T
	\end{equation}
	and the steering vector from RIS to the targets, $\mathbf{a}_{t,k}$ ,is defined as:
	\begin{equation}
		\begin{aligned}
			\mathbf{a}_{t,k} &= \begin{bmatrix} e^{-j2\pi \sigma_{1,k}/\lambda}, e^{-j2\pi \sigma_{2,k}/\lambda}, \ldots, e^{-j2\pi \sigma_{Q,k}/\lambda} \end{bmatrix}^T \\
			&= e^{-j2\pi [\boldsymbol{\Omega}]_{:,k}/\lambda}
		\end{aligned}
	\end{equation}
	where $\sigma_{q,k} = (\mathbf{p}_q^s - \mathbf{p}_1^s) \cdot \mathbf{q}_{t,k}$.Furthermore, $\boldsymbol{\Omega}$ is defined as:
	
	\begin{equation}
		\boldsymbol{\Omega} = \left( \mathbf{P}^s - \mathbf{1}_Q\otimes[\mathbf{P}]_1^s \right) \mathbf{Q}_t
		\label{eq:11}
	\end{equation}
	where $\mathbf{P}^s = \left[ \left( \mathbf{p}_1^s \right)^T, \left( \mathbf{p}_2^s \right)^T, \ldots, \left( \mathbf{p}_Q^s \right)^T \right]^T$ represents the position matrix of the RIS with $[\mathbf{P}]_i^s$ representing the $i$-th row of the matrix $\mathbf{P}^s$, and $\mathbf{Q}_t = \left[ \mathbf{q}_{t,1}, \mathbf{q}_{t,2}, \ldots, \mathbf{q}_{t,K} \right]$ denotes the direction matrix formed by the column-wise stacking of the $\mathbf{q}_{t,k}$.
	
	Subsequently, the echo signal reflected from the $k$-th target can be expressed as:
	\begin{equation}
		\mathbf{r}_k(t, \tau) = u_k(\tau) \boldsymbol{\zeta}_k \left[ \mathbf{G} \text{diag}\{\mathbf{w}\} \mathbf{a}_{t,k} \right]^T \mathbf{s}(t)
    \label{eq:EMVS12}
	\end{equation}
	where $\tau$ denotes the pulse index, $u_k(\tau)$ represents the reflection coefficient of the $k$-th target, $\mathbf{G} = \left[ \left[ \mathbf{G} \right]_1^T \left[ \mathbf{G} \right]_2^T \cdots \left[ \mathbf{G} \right]_M^T \right]^T$, and $\mathbf{s}(t) = \left[ s_1(t), \ldots, s_M(t) \right]^T$ represents the transmit signal vector.
	
	The received signal captures by receive EMVS array due to the target echoes, can be expressed as
	\begin{equation}
		\mathbf{y}(t, \tau) = \sum_{k=1}^K \mathbf{a}_{r,k} \otimes \left[ \mathbf{V}_{r,k} \boldsymbol{\Xi}_k \mathbf{r}_k(t, \tau) \right] + \mathbf{c}(t, \tau)
		\label{eq:EMVS13}
	\end{equation}
	where $\mathbf{V}_{r,k} \in \mathbb{C}^{6 \times 2}$ denotes the direction-dependent matrix of the $k$-th target spatial-polar response to the receive array, characterized by of $\theta_{r,k}$ and $\varphi_{r,k}$. $\boldsymbol{\Xi}_k \in \mathbb{C}^{2 \times 2}$ represents the scattering matrix of the $k$-th target, which fully represents its polarimetric transformation properties, and $\mathbf{c}(t, \tau)$ denotes the additive white Gaussian noise vector.Forthemore $\mathbf{a}_{r,k}$ is the steering vector at the receiver, denoted as
	\begin{equation}
		\begin{aligned}
			\mathbf{a}_{r,k} &= \begin{bmatrix} e^{-j2\pi \varsigma_{1,k}/\lambda}, e^{-j2\pi \varsigma_{2,k}/\lambda}, \ldots, e^{-j2\pi \varsigma_{N,k}/\lambda} \end{bmatrix}^T \\
			&= e^{-j2\pi [\boldsymbol{\Psi}]_{:,k}/\lambda}
		\end{aligned}
	\end{equation}
	where $\varsigma_{n,k} = \mathbf{p}_n^r \mathbf{q}_{r,k}$, and $\boldsymbol{\Psi}$ is defined as
	\begin{equation}
		\boldsymbol{\Psi} = \mathbf{P}^r \mathbf{Q}_r
		\label{eq:15}
	\end{equation}
	with $\mathbf{P}^r = \begin{bmatrix} \left( \mathbf{p}_1^r \right)^T, \left( \mathbf{p}_2^r \right)^T, \ldots, \left( \mathbf{p}_N^r \right)^T \end{bmatrix}^T$ representing the position matrix of the receive EMVS elements, and $\mathbf{Q}_r = \begin{bmatrix} \mathbf{q}_{r,1}, \mathbf{q}_{r,2}, \ldots, \mathbf{q}_{r,K} \end{bmatrix}$ denotes the direction matrix formed by the column-wise stacking of the $\mathbf{q}_{r,k}$. Combining \eqref{eq:EMVS12} and \eqref{eq:EMVS13}, the received signal can be rewritten as:
	\begin{equation}
		\begin{aligned}
			& \mathbf{y}(t, \tau) \\
			&= \sum_{k=1}^K u_k(\tau) \mathbf{a}_{r,k} \otimes \left[ \mathbf{V}_{r,k} \mathbf{g}_k \left[ \mathbf{G} \text{diag}\{\mathbf{w}\} \mathbf{a}_{t,k} \right]^T \mathbf{s}(t) \right] + \mathbf{c}(t, \tau) \\
			&= \sum_{k=1}^K (\mathbf{a}_{r,k} \otimes \mathbf{b}_k) u_k(\tau) \left[ \mathbf{G} \text{diag}\{\mathbf{w}\} \mathbf{a}_{t,k} \right]^T \mathbf{s}(t) + \mathbf{c}(t, \tau)
		\end{aligned}
	\end{equation}
	where $\mathbf{g}_k = \boldsymbol{\Xi}_k \boldsymbol{\zeta}_{k}$ and $\mathbf{b}_k = \mathbf{V}_{r,k} \mathbf{g}_k$ denote the polarization vector and the polarization response of the $k$-th target,respectively. Assume that the mutually orthogonal waveforms transmitted by the emitter elements satisfy
	\begin{equation}
		\int_{T_m} s_i(t) s_j^*(t) dt = \begin{cases} 1, & i = j \\ 0, & \text{else} \end{cases}
	\end{equation}
	where $T_m$ denotes the pulse duration. The received signal is then processed through matched filtering with the $m$-th transmitted waveform, yielding:
	\begin{equation}
		\begin{aligned}
			\mathbf{y}_m(\tau) &= \int \mathbf{y}(t, \tau) s_m^*(t) dt \\
			&= \sum_{k=1}^K (\mathbf{a}_{r,k} \otimes \mathbf{b}_k) u_k(\tau) \left[ \left[ \mathbf{G} \right]_m \text{diag}\{\mathbf{w}\} \mathbf{a}_{t,k} \right] + \mathbf{n}_m(\tau)
		\end{aligned}
	\end{equation}
	where $\mathbf{n}_m(\tau) = \int \mathbf{c}(t, \tau) s_m^*(t) dt \in \mathbb{C}^{6N \times 1}$. Stacking the $\mathbf{y}_m(\tau)$ row-wise yields a new vector $\mathbf{y}(\tau) = \begin{bmatrix} \mathbf{y}_1^T(\tau), \mathbf{y}_2^T(\tau), \ldots, \mathbf{y}_M^T(\tau) \end{bmatrix}^T \in \mathbb{C}^{6MN \times 1}$:
	
	\begin{equation}
		\begin{aligned}
			\mathbf{y}(\tau) &= \sum_{k=1}^K \left( \left[ \mathbf{G} \text{diag}\{\mathbf{w}\} \mathbf{a}_{t,k} \right] \otimes \mathbf{a}_{r,k} \otimes \mathbf{b}_k \right) u_k(\tau) + \mathbf{n}(\tau) \\
			&= \left[ \mathbf{G} \text{diag}\{\mathbf{w}\} \mathbf{A}_t \odot \mathbf{A}_r \odot \mathbf{B} \right] \mathbf{u}(\tau) + \mathbf{n}(\tau)
		\end{aligned}
		\label{eq:EMVS19}
	\end{equation}
	
	\noindent where $\mathbf{A}_t = \left[ \mathbf{a}_{t,1}, \mathbf{a}_{t,2}, \ldots, \mathbf{a}_{t,K} \right] \in \mathbb{C}^{Q \times K}$, $\mathbf{A}_r = \left[ \mathbf{a}_{r,1}, \mathbf{a}_{r,2}, \ldots, \mathbf{a}_{r,K} \right] \in \mathbb{C}^{N \times K}$, $\mathbf{B} = \left[ \mathbf{b}_1, \mathbf{b}_2, \ldots, \mathbf{b}_K \right] \in \mathbb{C}^{6 \times K}$, $\mathbf{u}(\tau) = \left[ u_1(\tau), u_2(\tau), \ldots, u_K(\tau) \right]^T \in \mathbb{C}^{K \times 1}$, and $\mathbf{n}(\tau) = \left[ \mathbf{n}_1^T(\tau), \ldots, \mathbf{n}_M^T(\tau) \right]^T \in \mathbb{C}^{6MN \times 1}$.

	After straightforward rearrangement, \eqref{eq:EMVS19} can be rewritten as:
	\begin{equation}
		\begin{aligned}
			\mathbf{y}(\tau) &= \left[ \mathbf{G} \text{diag}\{\mathbf{w}\} \mathbf{A}_t \odot \mathbf{A}_r \odot \mathbf{B} \right] \mathbf{u}(\tau) + \mathbf{n}(\tau) \\
			&= \left[ \mathbf{V}_t \odot \mathbf{A}_r \odot \mathbf{B} \right] \mathbf{u}(\tau) + \mathbf{n}(\tau)
		\end{aligned}
		\label{eq:20}
	\end{equation}
	where $\mathbf{V}_t = \mathbf{G} \text{diag}\{\mathbf{w}\} \mathbf{A}_t \in \mathbb{C}^{M \times K},\mathbf{C}_r=\mathbf{A}_r \odot \mathbf{B}\in \mathbb{C}^{6N \times K}$. Then, \eqref{eq:20} can be written in matrix format as
	\begin{equation}
		\begin{aligned}
			\mathbf{Y} &= \left[ \mathbf{G} \text{diag}\{\mathbf{w}\} \mathbf{A}_t \odot \mathbf{A}_r \odot \mathbf{B} \right] \mathbf{U} + \mathbf{N} \\
			&= \left[ \mathbf{V}_t \odot \mathbf{C}_r \right] \mathbf{U} + \mathbf{N}
		\end{aligned}
	\end{equation}
	where $\mathbf{Y} = \left[ \mathbf{y}(1), \mathbf{y}(2), \ldots, \mathbf{y}(L) \right] \in \mathbb{C}^{6MN \times L}$, $\mathbf{U} = \left[ \mathbf{u}(1), \mathbf{u}(2), \ldots, \mathbf{u}(L) \right] \in \mathbb{C}^{K \times L}$, and $\mathbf{N} = \left[ \mathbf{n}(1), \mathbf{n}(2), \ldots, \mathbf{n}(L) \right] \in \mathbb{C}^{6MN \times L}$.
	\section{The Proposed Algorithm}
    \label{sec:estimation}
	This section presents the algorithmic solution to the titular question: How arbitrary EMVS arrays achieve unambiguous RIS-aided localization when element spacing exceeds $\lambda/2$. The core challenge lies in resolving two coupled complexities: (i) RIS phase shift modulation introduces bilinear coupling between unknown target parameters and reconfigurable phase shifts, and (ii) arbitrary array geometries create spatial ambiguities when inter-element phase differences exceed $2\pi$. Our solution decouples these challenges through a two-stage framework that first separates RIS effects from target signatures via tensor decomposition, then exploits the EMVS rotational invariance to disambiguate spatial parameters.
	
	The proposed algorithm consists of two synergistic stages: (i) \textit{tensor-based parameter estimation with phase disambiguation}, which decomposes the received tensor to isolate the transmit-side information (DOD and RIS effects) from the receive-side information (DOA and polarization), followed by a novel disambiguation procedure that resolves $2\pi$ phase wrapping in arbitrary geometries; and (ii) \textit{SDP-relaxed RIS phase shift optimization}, which leverages estimated parameters to design phase shifts that simultaneously maximize SNR and suppress spatial ambiguities. This dual approach—algorithmic disambiguation combined with physical channel shaping —enables unambiguous localization beyond the traditional $\lambda/2$ constraint while maintaining near-optimal estimation accuracy.
	
	\subsection{Parallel Factor Analysis}
	The first step toward answering the beyond-$\lambda/2$ question is decoupling RIS phase shift effects from target spatial signatures—a prerequisite for applying phase disambiguation to arbitrary geometries. We employ PARAFAC decomposition, which uniquely factorizes the received tensor into matrices corresponding to different parameter spaces. Critically for our framework, PARAFAC naturally separates the transmit-side factor $\mathbf{V}_t$ (encoding both DOD and RIS modulation $\mathbf{G}\text{diag}\{\mathbf{w}\}\mathbf{A}_t$) from the receive-side factor $\mathbf{C}_r$ (encoding DOA and polarization). This separation is essential: Once $\mathbf{C}_r$ is isolated, we can apply phase disambiguation to resolve ambiguities arising from arbitrary receiver geometry without interference from RIS-induced phase shift modulation.
	By rearranging matrix $\mathbf{Y}$ into the form of a fourth-order tensor, we have
	\begin{equation}
		\mathcal{Y} = \sum_{k=1}^K \mathbf{v}_{t,k} \circ \mathbf{a}_{r,k} \circ \mathbf{b}_{r,k} \circ \mathbf{u}_k + \tilde{\mathcal{N}}
		\label{eq:22}
	\end{equation}
	where $\mathbf{u}_k \in \mathbb{C}^{L \times 1}$ denotes the $k$-th column of $\mathbf{U}$, $\mathbf{v}_{t,k} \in \mathbb{C}^{M \times 1}$ denotes the $k$-th column of $\mathbf{V}$, and $\tilde{\mathcal{N}}$ denotes the tensor form of noise matrix $\mathbf{N}$.
	
	According to the generalized tensorization operation of the PARAFAC model \cite{5}, the fourth-order tensor $\mathcal{Y}$ can be compressed into a third-order tensor $\mathcal{Z}$ by dividing the dimension indices $\mathbb{O} = \{1, 2, 3, 4\}$ into three non-overlapping order sets: $\mathbb{O}_1 = \{1\}$, $\mathbb{O}_2 = \{2, 3\}$, and $\mathbb{O}_3 = \{4\}$. Then, we obtain $\mathbf{g}_{1,k} = \mathbf{v}_{t,k}$, $\mathbf{g}_{2,k} = \mathbf{a}_{r,k} \circ \mathbf{b}_{r,k}$, and $\mathbf{g}_{3,k} = \mathbf{u}_k$. Let $\mathbf{V}_t = \begin{bmatrix} \mathbf{g}_{1,1}, \mathbf{g}_{1,2}, \ldots, \mathbf{g}_{1,K} \end{bmatrix}$, $\mathbf{C}_r = \begin{bmatrix} \mathbf{g}_{2,1}, \mathbf{g}_{2,2}, \ldots, \mathbf{g}_{2,K} \end{bmatrix}$, and $\mathbf{U} = \begin{bmatrix} \mathbf{g}_{3,1}, \mathbf{g}_{3,2}, \ldots, \mathbf{g}_{3,K} \end{bmatrix}$. Then, $\mathcal{Z}$ can be expressed as:
	\begin{equation}
		\mathcal{Z} = \mathcal{G}_{3,K \times 1} \mathbf{V}_{t \times 2} \mathbf{C}_{r \times 3} \mathbf{U} + \mathcal{N}
	\end{equation}
	where $\mathcal{G}_{3,K \times 1}$ is a $K \times K \times K$ identity tensor ,$\times j$ denotes the $j$-mode product of a tensor and a matrix. According to the definition of the PARAFAC decomposition, $\mathcal{Z}$ can be unfolded into three matrices as follows:
	
	\begin{equation}
		\mathbf{Z}_1 = [\mathcal{Z}]_{(1)}^T = (\mathbf{C}_r \odot \mathbf{U}) \mathbf{V}_t^T + \mathbf{N}_1
	\end{equation}
	
	\begin{equation}
		\mathbf{Z}_2 = [\mathcal{Z}]_{(2)}^T = (\mathbf{U} \odot \mathbf{V}_t) \mathbf{C}_r^T + \mathbf{N}_2
	\end{equation}
	
	\begin{equation}
		\mathbf{Z}_3 = [\mathcal{Z}]_{(3)}^T = (\mathbf{V}_t \odot \mathbf{C}_r) \mathbf{U}^T + \mathbf{N}_3
	\end{equation}
	
	By formulating the optimization problem as
	\begin{equation}
		\min_{\mathbf{V}_t, \mathbf{C}_r, \mathbf{U}} \|\mathcal{Z} - \mathcal{G}_{3,K \times 1} \mathbf{V}_{t \times 2} \mathbf{C}_{r \times 3} \mathbf{U}\|_F,
	\end{equation}
	the TALS algorithm \cite{5} can be adopted to estimate the factor matrices $\mathbf{V}_t$, $\mathbf{C}_r$, and $\mathbf{U}$ through the following iterative procedure.
	
	Fixing $\mathbf{C}_r$ and $\mathbf{U}$, the optimization problem becomes:
	\begin{equation}
		\min_{\mathbf{V}_t} \|\mathbf{Z}_1 - (\mathbf{C}_r \odot \mathbf{U}) \mathbf{V}_t^T\|_F.
	\end{equation}
	
	Therefore, the least-squares solution of $\mathbf{V}_t$ is given by
	\begin{equation}
		\hat{\mathbf{V}}_t^T = (\mathbf{C}_r \odot \mathbf{U})^\dagger \mathbf{Z}_1.
	\end{equation}
	
	Similarly, by fixing $\mathbf{V}_t$ and $\mathbf{U}$, the optimization problem becomes:
	\begin{equation}
		\min_{\mathbf{C}_r} \|\mathbf{Z}_2 - (\mathbf{U} \odot \mathbf{V}_t) \mathbf{C}_r^T\|_F.
	\end{equation}
	
	\noindent The least-squares solution of $\mathbf{C}_r$ is then given by
	\begin{equation}
		\hat{\mathbf{C}}_r^T = (\mathbf{U} \odot \mathbf{V}_t)^\dagger \mathbf{Z}_2.
	\end{equation}
	
	Finally, by fixing $\mathbf{V}_t$ and $\mathbf{C}_r$, the optimization problem becomes:
	\begin{equation}
		\min_{\mathbf{U}} \|\mathbf{Z}_3 - (\mathbf{V}_t \odot \mathbf{C}_r) \mathbf{U}^T\|_F.
	\end{equation}
	
	The least-squares solution of $\mathbf{U}$ is given by
	\begin{equation}
		\hat{\mathbf{U}}^T = (\mathbf{V}_t \odot \mathbf{C}_r)^\dagger \mathbf{Z}_3.
	\end{equation}
	
	Assuming that the factor matrices $\mathbf{V}_t$, $\mathbf{C}_r$, and $\mathbf{U}$ are all full column rank with Kruskal ranks $k_{\mathbf{V}_t}$, $k_{\mathbf{C}_r}$, and $k_{\mathbf{U}}$, respectively, and that they satisfy the Kruskal condition:
	\begin{equation}
		k_{\mathbf{V}_t} + k_{\mathbf{C}_r} + k_{\mathbf{U}} \geq 2K + 2,
	\end{equation}
	then $\mathbf{V}_t$, $\mathbf{C}_r$, and $\mathbf{U}$ are essentially unique up to column permutation and scaling transformations. That is, the estimated factor matrices satisfy:
	\begin{equation}
		\hat{\mathbf{V}}_t = \mathbf{V}_t  \boldsymbol{\Pi} \boldsymbol{\Lambda}_1 + \mathbf{F}_1
	\end{equation}
	\begin{equation}
		\hat{\mathbf{C}}_r = \mathbf{C}_r  \boldsymbol{\Pi} \boldsymbol{\Lambda}_2 + \mathbf{F}_2
	\end{equation}
	\begin{equation}
		\hat{\mathbf{U}} = \mathbf{U} \boldsymbol{\Pi} \boldsymbol{\Lambda}_3 + \mathbf{F}_3
		\label{eq:37}
	\end{equation}
	where $\hat{\mathbf{V}}_t$, $\hat{\mathbf{C}}_r$, and $\hat{\mathbf{U}}$ are matrices estimated via PARAFAC decomposition exhibiting column ambiguity and scaling ambiguity. $\boldsymbol{\Pi}$ is the column permutation matrix, $\boldsymbol{\Lambda}_1$, $\boldsymbol{\Lambda}_2$, and $\boldsymbol{\Lambda}_3$ are scaling ambiguity matrices with non-zero diagonal elements, while $\mathbf{F}_1$, $\mathbf{F}_2$, and $\mathbf{F}_3$ represent fitting errors.
	
	\subsection{2-D DOD Estimation}
	Having obtained the factor matrix $\hat{\mathbf{V}}_t$ from PARAFAC decomposition, we now proceed to extract the 2-D DOD information. Note that $\mathbf{V}_t$ contains the coupled information of the transmit array geometry and the RIS phase shifts. By exploiting the known RIS configuration matrix $\mathbf{G}$ and phase shift vector $\mathbf{w}$, we can decouple this information to recover the transmit steering matrix $\mathbf{A}_t$.
	By resolving the scaling and permutation ambiguities in matrix $\hat{\mathbf{V}}_t$ to obtain $\hat{\mathbf{V}}_t'$ and combining it with \eqref{eq:20}, we have
	\begin{equation}
		\hat{\mathbf{A}}_t =  \left( \mathbf{G} \text{diag}\{\mathbf{w}\} \right)^\dagger \hat{\mathbf{V}}_t'
		\label{eq:38}
	\end{equation}
	
	With $\mathbf{A}_t = \left[ \mathbf{a}_{t1}, \mathbf{a}_{t2}, \ldots, \mathbf{a}_{tK} \right]$, an estimated matrix related to the 2-D DOD can be computed as
	\begin{equation}
		\hat{\boldsymbol{\Omega}} = \frac{\text{angle}(\hat{\mathbf{A}}_t) \lambda}{2\pi}
	\end{equation}
	
	Applying the least-squares(LS) principle to  \eqref{eq:11}, we have
	\begin{equation}
		\hat{\mathbf{Q}}_t =  \left( \mathbf{P}^s -  \mathbf{1}_Q\otimes\mathbf{P}_1^s \right)^\dagger \hat{\boldsymbol{\Omega}}
		\label{eq:40}
	\end{equation}
	
	Therefore, the 2-D DOD estimates can be obtained as
	\begin{equation}
		\begin{aligned}
			\hat{\theta}_{t,k} &= \arctan \left( \frac{\hat{\mathbf{Q}}_t(2,k)}{\hat{\mathbf{Q}}_t(1,k)} \right) \\
			\hat{\varphi}_{t,k} &= \arcsin \left( \sqrt{\hat{\mathbf{Q}}_t(1,k)^2 + \hat{\mathbf{Q}}_t(2,k)^2} \right)
		\end{aligned}
	\end{equation}
	{where $\hat{\mathbf{Q}}_t(m,k)$ is the element in row $m$ and column $k$ of matrix $\hat{\mathbf{Q}}_t$.}
	\subsection{2-D DOA Coarse Estimation}
	With the DOD parameters successfully estimated, we now turn our attention to the receiving side to estimate the 2-D DOA parameters. Unlike conventional scalar sensor arrays, EMVS arrays measure both the electric and magnetic field components, providing rich polarimetric information. We exploit the rotational invariance property inherent in the six-component EMVS structure to obtain an initial coarse estimate of the DOA, which serves as a foundation for subsequent refinement.
	First, we construct the transformation matrices as
	\begin{equation}
		\mathbf{H}_1 = \mathbf{I}_N \otimes \mathbf{i}_{6,1}
	\end{equation}
	
	\begin{equation}
		\mathbf{H}_2 = \mathbf{I}_N \otimes \mathbf{i}_{6,2}
	\end{equation}
	
	Similarly, we have $\mathbf{H}_3$, $\mathbf{H}_4$, $\mathbf{H}_5$, and $\mathbf{H}_6$. Ignoring the error terms yields
	\begin{equation}
		\mathbf{H}_r \hat{\mathbf{C}}_r \approx \mathbf{H}_r \mathbf{C}_r \boldsymbol{\Pi} \boldsymbol{\Lambda}_2
	\end{equation}
	where $\hat{\mathbf{C}}_r = \hat{\mathbf{A}}_r \odot \hat{\mathbf{B}}_r$, and $\mathbf{H}_r = \mathbf{I}_N \otimes \mathbf{i}_{6,r}$ for $r = 1, 2, \ldots, 6$. By exploiting the rotational invariance property, we obtain
	\begin{equation}
		\mathbf{H}_1 \hat{\mathbf{C}}_r \boldsymbol{\Phi}_q = \mathbf{H}_q \hat{\mathbf{C}}_r
		\label{eq:45}
	\end{equation}
	where $\boldsymbol{\Phi}_q = \text{diag} \left\{ \frac{\hat{b}_1(q)}{\hat{b}_1(1)}, \frac{\hat{b}_2(q)}{\hat{b}_2(1)}, \ldots, \frac{\hat{b}_K(q)}{\hat{b}_K(1)} \right\}$ for $q = 1, 2, 3, 4, 5, 6$.
	
	Let $\hat{\mathbf{e}}_k = \begin{bmatrix} 1, \frac{\hat{b}_k(2)}{\hat{b}_k(1)}, \frac{\hat{b}_k(3)}{\hat{b}_k(1)} \end{bmatrix}^T$ and $\hat{\mathbf{m}}_k = \begin{bmatrix} \frac{\hat{b}_k(4)}{\hat{b}_k(1)}, \frac{\hat{b}_k(5)}{\hat{b}_k(1)}, \frac{\hat{b}_k(6)}{\hat{b}_k(1)} \end{bmatrix}^T$. The estimate $\hat{\mathbf{q}}_{r,k}$ can be obtained via the vector cross-product as follows:
	\begin{equation}
		\hat{\mathbf{q}}_{r,k} = \frac{\hat{\mathbf{e}}_k}{\|\hat{\mathbf{e}}_k\|_F} \circledast \frac{\hat{\mathbf{m}}_k^*}{\|\hat{\mathbf{m}}_k\|_F}
		\label{eq:46}
	\end{equation}
	
	Subsequently, the coarse 2-D DOA estimates are obtained from \eqref{eq:46} as
	\begin{equation}
		\hat{\theta}_{r,k} = \arctan \left( \frac{\hat{\mathbf{q}}_{r,k}(2)}{\hat{\mathbf{q}}_{r,k}(1)} \right)
	\end{equation}
	\begin{equation}
		\hat{\varphi}_{r,k} = \arcsin \left( \sqrt{\hat{\mathbf{q}}_{r,k}(1)^2 + \hat{\mathbf{q}}_{r,k}(2)^2} \right)
	\end{equation}
	{where $\hat{\mathbf{q}}_{r,k}(m)$  refers to the element at index $m$  in vector $\hat{\mathbf{q}}_{r,k}$.}
	\subsection{Ambiguity-resolved 2-D DOA Estimation}
	While the VCP-based coarse estimation provides an initial DOA estimate, it does not fully exploit the phase information available in the steering matrix. To achieve higher estimation accuracy, particularly in scenarios with large aperture arrays where the inter-element spacing exceeds half a wavelength, we need to resolve the phase ambiguity and refine the DOA estimates using the complete phase information.
	The aforementioned method fails to fully leverage the phase information in the steering matrix, the utilization of which for DOA estimation can yield higher accuracy. However, phase ambiguity arises when extracting the phase via the angle function, which only resolves angles within $\left( -\pi, \pi \right]$. When the inter-element spacing exceeds half a wavelength, the phase difference between adjacent elements exceeds $2\pi$, thereby causing phase ambiguity.
	
	The phase components of steering matrix $\mathbf{A}_r$ can be represented as:
	\begin{equation}
		\mathbf{T} = \begin{bmatrix}
			2\pi \varsigma_{1,1}/\lambda & 2\pi \varsigma_{1,2}/\lambda & \cdots & 2\pi \varsigma_{1,K}/\lambda \\
			2\pi \varsigma_{2,1}/\lambda & 2\pi \varsigma_{2,2}/\lambda & \cdots & 2\pi \varsigma_{2,K}/\lambda \\
			\vdots & \vdots & \ddots & \vdots \\
			2\pi \varsigma_{N,1}/\lambda & 2\pi \varsigma_{N,2}/\lambda & \cdots & 2\pi \varsigma_{N,K}/\lambda
		\end{bmatrix}
		\label{eq:49}
	\end{equation}
	
	Up to this point, $\hat{\mathbf{C}}_r$ has been estimated; however, $\mathbf{A}_r$ remains unknown, but is required in the phase disambiguation process. Therefore, appropriate processing of $\hat{\mathbf{C}}_r$ is necessary. First, we perform column permutation and phase disambiguation on $\hat{\mathbf{C}}_r$ to obtain $\hat{\mathbf{C}}_r'$. Note that $\hat{\mathbf{C}}_r' = \hat{\mathbf{A}}_r \odot \hat{\mathbf{B}}_r$, and $\hat{\mathbf{b}}_k(6)$ is real-valued, yielding
	\begin{equation}
		\hat{\mathbf{D}}' = \mathbf{H}_6 \hat{\mathbf{C}}_r' = \begin{bmatrix} \hat{\mathbf{A}}_r(1,1) \hat{\mathbf{b}}_1(6) & \cdots & \hat{\mathbf{A}}_r(1,K) \hat{\mathbf{b}}_K(6) \\ \vdots & \ddots & \vdots \\ \hat{\mathbf{A}}_r(N,1) \hat{\mathbf{b}}_1(6) & \cdots & \hat{\mathbf{A}}_r(N,K) \hat{\mathbf{b}}_K(6) \end{bmatrix}
		\label{eq:50}
	\end{equation}
	
	Extracting the phase yields
	\begin{equation}
		\hat{\boldsymbol{\mathbf{T}}}' = \text{angle} \left\{ \hat{\mathbf{D}}' \right\}
		\label{eq:51}
	\end{equation}
	
	Based on the coarse estimates, we construct the direction matrix $\hat{\mathbf{Q}}_r = \left[ \hat{\mathbf{q}}_{r,1}, \hat{\mathbf{q}}_{r,2}, \ldots, \hat{\mathbf{q}}_{r,K} \right]$. Together with \eqref{eq:15} and \eqref{eq:49}, we have
	\begin{equation}
		\hat{\boldsymbol{\mathbf{T}}} = \frac{2\pi}{\lambda} \mathbf{P}^{r} \hat{\mathbf{Q}}_r
		\label{eq:52}
	\end{equation}
	
	For the steering vector, if the inter-element spacing exceeds half a wavelength, phase ambiguity occurs, manifested as a $2\pi$ phase difference. That is, there exists:
	\begin{equation}
		\hat{\boldsymbol{\mathbf{T}}}' - \hat{\boldsymbol{\mathbf{T}}} = 2\pi \mathbf{R}
	\end{equation}
	where $\mathbf{R}$ is an $N \times K$ integer matrix. Due to phase extraction constraints, for $e^{j\theta}$, the value remains identical for all $\theta$ modulo $2\pi$. In fact, when the phase exceeds $2\pi$, to obtain $\mathbf{R}$, we perform
	\begin{equation}
		\hat{\mathbf{R}} \triangleq \text{round} \left\{ \frac{\hat{\boldsymbol{\mathbf{T}}}' - \hat{\boldsymbol{\mathbf{T}}}}{2\pi} \right\}
		\label{eq:54}
	\end{equation}
	
	After phase compensation, the unambiguous phase matrix can be expressed as
	\begin{equation}
		\overline{\boldsymbol{\mathbf{T}}} = \hat{\boldsymbol{\mathbf{T}}}' + 2\pi \hat{\mathbf{R}}
		\label{eq:55}
	\end{equation}
	and then, the refined direction matrix can be obtained as
	\begin{equation}
		\overline{\mathbf{Q}}_r = \frac{\lambda}{2\pi} \left( \mathbf{P}^r \right)^\dagger \overline{\boldsymbol{\mathbf{T}}}.
		\label{eq:56}
	\end{equation}
	
	The refined 2-D DOA estimates can then be derived as
	\begin{equation}
		\begin{aligned}
			\overline{\theta}_{r,k} &= \arctan \left( \frac{\overline{\mathbf{Q}}_r(2,k)}{\overline{\mathbf{Q}}_r(1,k)} \right) \\
			\overline{\varphi}_{r,k} &= \arcsin \left( \sqrt{\overline{\mathbf{Q}}_r(1,k)^2 + \overline{\mathbf{Q}}_r(2,k)^2} \right)
		\end{aligned}
		\label{eq:57}
	\end{equation}

	\subsection{Polarization Parameter Estimation}
	Beyond spatial information, the polarization state of the scattered signal provides an additional dimension for target characterization. This is particularly valuable in RIS-aided systems where multiple targets may share similar spatial angles. By estimating polarization parameters $\zeta_k$ and $\varrho_k$, we can uniquely identify and distinguish different targets.
	Estimating the polarization parameters is crucial as it allows us to distinguish different targets located in the same direction. First, utilizing the unambiguous direction matrix $\overline{\mathbf{Q}}_r$ from \eqref{eq:56}, we obtain the corresponding steering matrix $\overline{\mathbf{A}}_r$ as follows:
	\begin{equation}
		\overline{\mathbf{A}}_r = \begin{bmatrix}
			e^{-j2\pi \overline{\varsigma}_{1,1}/\lambda} & e^{-j2\pi \overline{\varsigma}_{1,2}/\lambda} & \cdots & e^{-j2\pi \overline{\varsigma}_{1,K}/\lambda} \\
			e^{-j2\pi \overline{\varsigma}_{2,1}/\lambda} & e^{-j2\pi \overline{\varsigma}_{2,2}/\lambda} & \cdots & e^{-j2\pi \overline{\varsigma}_{2,K}/\lambda} \\
			\vdots & \vdots & \ddots & \vdots \\
			e^{-j2\pi \overline{\varsigma}_{N,1}/\lambda} & e^{-j2\pi \overline{\varsigma}_{N,2}/\lambda} & \cdots & e^{-j2\pi \overline{\varsigma}_{N,K}/\lambda}
		\end{bmatrix}
		\label{eq:58}
	\end{equation}
	
	Then, we construct two auxiliary matrices as
	\begin{equation}
		\mathbf{J} = \text{diag}(\mathbf{i}_{N,1}) \otimes \mathbf{I}_6 \in \mathbb{C}^{6N \times 6N}
		\label{eq:59}
	\end{equation}
	\begin{equation}
		\mathbf{D} =
		\begin{bmatrix}
			\frac{1}{\overline{\mathbf{A}}_r (1,1)} & \frac{1}{\overline{\mathbf{A}}_r (1,2)} & \cdots & \frac{1}{\overline{\mathbf{A}}_r (1,K)} \\
			\frac{1}{\overline{\mathbf{A}}_r (1,1)} & \frac{1}{\overline{\mathbf{A}}_r (1,2)} & \cdots & \frac{1}{\overline{\mathbf{A}}_r (1,K)} \\
			\vdots & \vdots & \ddots & \vdots \\
			\frac{1}{\overline{\mathbf{A}}_r (1,1)} & \frac{1}{\overline{\mathbf{A}}_r (1,2)} & \cdots & \frac{1}{\overline{\mathbf{A}}_r (1,K)}
		\end{bmatrix}
		\in \mathbb{C}^{6\times K}
		\label{eq:60}
	\end{equation}
	
	We have
	\begin{equation}
		\overline{\mathbf{B}} = \mathbf{J} \hat{\mathbf{C}}_r' \oplus \mathbf{D}
		\label{eq:61}
	\end{equation}
	
	From this, a polarization-dependent vector can be calculated as
	\begin{equation}
		\overline{\mathbf{g}}_k =  \overline{\mathbf{V}}^\dagger \overline{\mathbf{b}}_k
	\end{equation}
	where $\overline{\mathbf{V}}$ represents the estimated direction-dependent matrix derived based on \eqref{eq:3} with the estimates of the 2-D DOA in \eqref{eq:57}, and $\overline{\mathbf{b}}_k$ is the $k$-th column of $\overline{\mathbf{B}}$. Consequently, the 2-D polarization angles at the receiving EMVS for the $k$-th target can be calculated as
	\begin{equation}
		\begin{aligned}
			\overline{\zeta}_k &= \arctan \left(| \frac{\overline{\mathbf{g}}_k(1)}{\overline{\mathbf{g}}_k(2)} | \right) \\
			\overline{\varrho}_k &= \text{angle} \left( \frac{\overline{\mathbf{g}}_k(1)}{\overline{\mathbf{g}}_k(2)} \right)
		\end{aligned}
	\end{equation}

	\subsection{Design of RIS Phase Shifts}
	Up to this point, we have focused on extracting target parameters from the received signals for a given RIS configuration. However, a critical advantage of RIS-aided radar systems lies in their ability to intelligently configure the RIS phase shifts to enhance system performance. Specifically, by optimizing the RIS phase shifts based on the estimated target parameters, we can coherently combine the scattered signals at the receiver, thereby significantly improving the SNR and detection capability.
	
	The key insight is that the RIS acts as a \textit{passive beamformer} that can focus the reflected energy toward the receiver without requiring additional power consumption. Unlike traditional active relays, the RIS achieves this through passive phase modulation, making it energy-efficient and cost-effective. The optimization objective is to design the phase shift vector $\mathbf{w}$ such that the received signal power is maximized.
	Optimizing the RIS phase shifts can enhance the SNR at the receiver. Considering that the RIS only alters the signal phase, the optimization problem for maximizing the received signal power is formulated as follows:
	\begin{equation}
		\begin{aligned}
			& \max_{\mathbf{w}} && \| [\mathbf{G}\text{diag}\{\mathbf{w}\} \mathbf{A}_t \odot \mathbf{C}] \mathbf{U} \|_F^2 \\
			& \text{s.t.} && |\mathbf{w}_n| = 1, \quad n = 1, 2, \cdots, Q
		\end{aligned}
		\label{eq:64}
	\end{equation}
	where $\mathbf{C} = \mathbf{A}_r \odot \mathbf{B} \in \mathbb{C}^{6N \times K}$. From the properties of the Frobenius norm, we deduce that
	\begin{equation}
		\scalebox{0.8}{$
			\begin{aligned}
				\| \left[ \mathbf{G} \text{diag}\{\mathbf{w}\} \mathbf{A}_t \odot \mathbf{C}\right] \mathbf{U} \|_F^2 = \text{tr} \left[ \mathbf{U}^H \left( \mathbf{V}(\mathbf{w}) \odot \mathbf{C} \right)^H \left( \mathbf{V}(\mathbf{w}) \odot \mathbf{C} \right) \mathbf{U} \right]
			\end{aligned}
			$}
	\end{equation}
	where $\mathbf{V}(\mathbf{w}) = \mathbf{G} \text{diag}\{\mathbf{w}\} \mathbf{A}_t$. Subsequently, defining $\mathbf{R}(\mathbf{w}) = \left( \mathbf{V}(\mathbf{w}) \odot \mathbf{C} \right)^H \left( \mathbf{V}(\mathbf{w}) \odot \mathbf{C} \right)$, we have
	\begin{equation}
		\| \left[ \mathbf{G} \text{diag}\{\mathbf{w}\} \mathbf{A}_t \odot \mathbf{C}\right] \mathbf{U} \|_F^2  = \text{tr} \left[ \mathbf{U}^H \mathbf{R}(\mathbf{w}) \mathbf{U} \right]
		\label{eq:66}
	\end{equation}
	Through further deduction (See Appendix), \eqref{eq:66} can be converted into:
	\begin{equation}
		\text{tr} \left[ \mathbf{U}^H \mathbf{R}(\mathbf{w}) \mathbf{U} \right] = \mathbf{w}^H \boldsymbol{\Phi} \mathbf{w}
		\label{eq:67}
	\end{equation}
	where
	\begin{equation}
		\boldsymbol{\Phi} = \sum_{i=1}^K \sum_{j=1}^K \left[ \mathbf{U}^H \mathbf{U} \right]_{ij} \left[ \mathbf{C}^H \mathbf{C} \right]_{ij} \left[ \left( \mathbf{G}^H \mathbf{G} \right) \oplus \left( \mathbf{a}_{t,i}^* \mathbf{a}_{t,j}^T \right) \right]
		\label{eq:68}
	\end{equation}
	with $\mathbf{a}_{t,i}$ and $\mathbf{a}_{t,j}$ denoting the $i$-th and $j$-th columns of matrix $\mathbf{A}_t$, respectively.
	
	Subsequently, the optimization problem in \eqref{eq:64} is transformed into:
	\begin{equation}
		\max_{\mathbf{w}} \mathbf{w}^H \boldsymbol{\Phi} \mathbf{w} \quad \text{s.t.} \quad |\mathbf{w}_n| = 1, n = 1, 2, \cdots, Q
		\label{eq:69}
	\end{equation}
	
	Let $\mathbf{X} = \mathbf{w} \mathbf{w}^H$, where $\mathbf{X}$ is a Hermitian positive-semidefinite matrix of rank one. The constraint that all diagonal elements equal unity implies that $\mathbf{X}$ is a phase-only weighting matrix. The objective function in \eqref{eq:69} can be rewritten as:
	\begin{equation}
		\mathbf{w}^H \boldsymbol{\Phi} \mathbf{w} = \text{tr} \left\{ \boldsymbol{\Phi} \mathbf{w} \mathbf{w}^H \right\} = \text{tr} \left\{ \boldsymbol{\Phi} \mathbf{X} \right\}
	\end{equation}
	such that, \eqref{eq:69} can be reformulated as:
	\begin{equation}
		\begin{aligned}
			& \max_{\mathbf{X}} \quad \mathrm{tr}\{\mathbf{\Phi} \mathbf{X}\} \\
			& \text{s.t.} \quad \mathrm{rank}(\mathbf{X}) = 1, \\
			& \qquad \quad \mathbf{X}_{nn} = 1, \quad n = 1, 2, \cdots, Q, \\
			& \qquad \quad \mathbf{X} \succcurlyeq 0
		\end{aligned}
		\label{eq:71}
	\end{equation}
	
	In \eqref{eq:71}, except for the rank constraint $\text{rank}(\mathbf{X}) = 1$, both the objective function and all constraints regarding $\mathbf{X}$ are convex. The constraint $\text{rank}(\mathbf{X}) = 1$ can be yield removed to
	\begin{equation}
		\begin{aligned}
			& \max_{\mathbf{X}} \quad \mathrm{tr}\{\mathbf{\Phi} \mathbf{X}\} \\
			& \text{s.t.} \quad \mathbf{X} \succcurlyeq 0, \\
			& \qquad \quad \mathbf{X}_{nn} = 1, \quad n = 1, \ldots, Q
		\end{aligned}
		\label{eq:72}
	\end{equation}
	
	It can be observed that \eqref{eq:72} constitutes a convex optimization problem. However, a critical issue arises: the globally optimal solution $\mathbf{X}'$ obtained via the interior-point algorithm may not be of rank 1, and it has to be transformed into a feasible solution $\mathbf{w}'$ for \eqref{eq:64}. To address this issue, we perform Cholesky decomposition on $\mathbf{X}'$ and employ Gaussian randomization to extract an approximate feasible solution $\mathbf{w}'$. To proceed, let
	\begin{equation}
		\mathbf{X}' = \mathbf{V}^H  \mathbf{V} \quad \left( \mathbf{V} \in \mathbb{C}^{r \times Q}, r = \text{rank}(\mathbf{X}') \right)
		\label{eq:73}
	\end{equation}
	where the $l$-th $(l=1,2,\ldots,L)$ random vector of size $r \times 1$ is generated from
	\begin{equation}
		\mathbf{z}_l \sim \mathcal{CN}(0, \mathbf{I}_r), \mathbf{z}_l \in \mathbb{C}^{r \times 1}
		\label{eq:74}
	\end{equation}
	and $\mathcal{CN}(0, \mathbf{I}_r)$ denotes the complex Gaussian distribution with zero mean and covariance matrix $\mathbf{I}_r$. The candidate vector is computed as
	\begin{equation}
		\mathbf{v}_l = \mathbf{V}^H \mathbf{z}_l \in \mathbb{C}^{Q \times 1}
		\label{eq:75}
	\end{equation}
	
	Then, $\mathbf{v}_l$ is mapped to the feasible solution space of \eqref{eq:64} as follows
	\begin{equation}
		\mathbf{w}_l^{(n)} = e^{j \times \text{angle}(\mathbf{v}_l^{(n)})}
		\label{eq:76}
	\end{equation}
	where $\mathbf{v}_l^{(n)}$ denotes the $n$-th element of the candidate vector $\mathbf{v}_l$. Finally, select the solution that maximizes the output power from the generated $L$ feasible solutions as the approximately optimal solution to \eqref{eq:64}, shown as
	\begin{equation}
		\begin{aligned}
			l' &= \arg\max_{l=1,2,\cdots,L} \mathbf{w}_l^H \mathbf{\Phi} \mathbf{w}_l\\
			\mathbf{w}' &= \mathbf{w}_{l'}
		\end{aligned}
	\end{equation}
 
	\section{Algorithmic Analysis}
	\label{sec:analysis}
	\subsection{Maximum Number of Identifiable Targets}
	Kruskal's theorem \cite{5} establishes the upper limit on the number of identifiable targets, $K_{max}$, by defining the uniqueness condition of the PARAFAC decomposition. Given that $\mathbf{V}_t \in \mathbb{R}^{M \times K}$, $\mathbf{C}_r \in \mathbb{R}^{6N \times K}$, and $\mathbf{U} \in \mathbb{R}^{L \times K}$ are full column rank matrices, the Kruskal ranks satisfy $\text{rank}\{\mathbf{V}_t\} = \min\{M, K\}$, $\text{rank}\{\mathbf{C}_r\} = \min\{6N, K\}$, and $\text{rank}\{\mathbf{U}\} = \min\{L, K\}$. Thus, it follows that \cite{32}
	\begin{equation}
		K_{max} = M + 6N - 2
		\label{eq:78}
	\end{equation}
	
	Additionally, $K$ is constrained by the rotational invariance property in \eqref{eq:45}; specifically, $K$ must satisfy
	\begin{equation}
		K_{max} \leq N
		\label{eq:79}
	\end{equation}
	
	Thus, combining \eqref{eq:78} and \eqref{eq:79}, the maximum number of targets that can be identified is:
	\begin{equation}
		K_{max} = N
	\end{equation}

	\subsection{Computational Complexity}
	Here, we analyze the computational complexity of the proposed method by focusing on the main components, including PARAFAC decomposition, RIS phase optimization, and Gaussian randomization. The dominant complexity of the PARAFAC-based estimator stems from the truncated eigenvalue decomposition during initialization, which involves a complexity of $\mathcal{O}\left( N^3 + 6^3 M^3 + L^3 \right)$ for the $N \times 6M \times L$-dimensional tensor, where $\mathcal{O}(\cdot)$ denotes the order of complexity. Regarding RIS phase optimization, computing the matrix $\boldsymbol{\Phi}$ has a complexity of $\mathcal{O}\left( K^2Q^2 + MQ^2 + 6NK^2 + K^2L \right)$, and the complexity of solving the SDP problem depends on the number of iterations, with the worst-case complexity being $\mathcal{O}\left( Q^4 \right)$. The complexity of Gaussian randomization is dominated by Cholesky decomposition, with a complexity of $\mathcal{O}\left( Q^3 + LQ^2 \right)$.
	
	Therefore, the total computational complexity of the proposed method is about \(\mathcal{O}\left( N^3 + 6^3 M^3 + L^3 + (K^2 + M + L)Q^2 + (6N + L)K^2  \right. \)
	\(\left.+ Q^4 + Q^3 \right)\).

	\subsection{Deterministic Cram\'er-Rao Bound}
	Here, we derive the Cram\'er-Rao bound (CRB) for the considered model as a benchmark for performance evaluation.
	
	First, define the unknown parameter vector \(\boldsymbol{\Theta} = \begin{bmatrix} \boldsymbol{\theta}_t^T, \boldsymbol{\theta}_r^T, \boldsymbol{\varphi}_t^T, \boldsymbol{\varphi}_r^T, \boldsymbol{\zeta}^T, \boldsymbol{\varrho}^T \end{bmatrix}^T\), where $\boldsymbol{\theta}_t = \left[ \theta_{t,1}, \theta_{t,2}, \ldots, \theta_{t,K} \right]^T$, $\boldsymbol{\theta}_r = \left[ \theta_{r,1}, \theta_{r,2}, \ldots, \theta_{r,K} \right]^T$, $\boldsymbol{\varphi}_r = \left[ \varphi_{r,1}, \varphi_{r,2}, \ldots, \varphi_{r,K} \right]^T$, $\boldsymbol{\varphi}_t = \left[ \varphi_{t,1}, \varphi_{t,2}, \ldots, \varphi_{t,K} \right]^T$, $\boldsymbol{\zeta} = \left[ \zeta_1, \zeta_2, \ldots, \zeta_K \right]^T$, and $\boldsymbol{\varrho} = \left[ \varrho_1, \varrho_2, \ldots, \varrho_K \right]^T$. According to \cite{33}, the $(m,n)$-th entry of the Fisher matrix for $\boldsymbol{\Theta}$ is given by
	\begin{equation}
		\text{FIM}_{(m,n)} = \frac{2L}{\sigma^2} \text{Re} \left\{ \frac{\partial \mathbf{F}^H}{\partial \boldsymbol{\Theta}_m} \mathbf{P}_\mathbf{F}^\perp \frac{\partial \mathbf{F}}{\partial \boldsymbol{\Theta}_n} \mathbf{R}_U \right\}
	\end{equation}
	where $\sigma^2$ denotes the noise power, $\mathbf{F} = \mathbf{V}_t \odot \mathbf{A}_r \odot \mathbf{B}$, $\mathbf{P}_\mathbf{F}^\perp = \mathbf{I}_{6MN} - \mathbf{F} (\mathbf{F}^H \mathbf{F})^{-1} \mathbf{F}^H$, and $\mathbf{R}_U = \frac{1}{L} \mathbf{U}^H \mathbf{U}$.
	
	Further, define: \(\tilde{\mathbf{F}} = \left[ \mathbf{F}_{\theta_t}, \mathbf{F}_{\theta_r}, \mathbf{F}_{\varphi_t}, \mathbf{F}_{\varphi_r}, \mathbf{F}_{\zeta}, \mathbf{F}_{\varrho} \right]\) with
	\scalebox{0.7}{\(\mathbf{F}_{\theta_r} = \begin{bmatrix} \dfrac{\partial \mathbf{f}_1}{\partial \theta_{r,1}}, & \dfrac{\partial \mathbf{f}_2}{\partial \theta_{r,2}}, & \cdots, & \dfrac{\partial \mathbf{f}_K}{\partial \theta_{r,K}} \end{bmatrix}\)},
	\scalebox{0.7}{\(\mathbf{F}_{\theta_t} = \begin{bmatrix} \dfrac{\partial \mathbf{f}_1}{\partial \theta_{t,1}}, & \dfrac{\partial \mathbf{f}_2}{\partial \theta_{t,2}}, & \cdots, & \dfrac{\partial \mathbf{f}_K}{\partial \theta_{t,K}} \end{bmatrix}\)},
	\scalebox{0.7}{\(\mathbf{F}_{\varphi_r} = \begin{bmatrix} \dfrac{\partial \mathbf{f}_1}{\partial \varphi_{r,1}}, & \dfrac{\partial \mathbf{f}_2}{\partial \varphi_{r,2}}, & \cdots, & \dfrac{\partial \mathbf{f}_K}{\partial \varphi_{r,K}} \end{bmatrix}\)},
	\scalebox{0.7}{\(\mathbf{F}_{\varphi_t} = \begin{bmatrix} \dfrac{\partial \mathbf{f}_1}{\partial \varphi_{t,1}}, & \dfrac{\partial \mathbf{f}_2}{\partial \varphi_{t,2}}, & \cdots, & \dfrac{\partial \mathbf{f}_K}{\partial \varphi_{t,K}} \end{bmatrix}\)},
	\scalebox{0.7}{\(\mathbf{F}_{\zeta} = \begin{bmatrix} \dfrac{\partial \mathbf{f}_1}{\partial \zeta_1}, & \dfrac{\partial \mathbf{f}_2}{\partial \zeta_2}, & \cdots, & \dfrac{\partial \mathbf{f}_K}{\partial \zeta_K} \end{bmatrix}\)},
	\scalebox{0.7}{\(\mathbf{F}_{\varrho} = \begin{bmatrix} \dfrac{\partial \mathbf{f}_1}{\partial \varrho_1}, & \dfrac{\partial \mathbf{f}_2}{\partial \varrho_2}, & \cdots, & \dfrac{\partial \mathbf{f}_K}{\partial \varrho_K} \end{bmatrix}\)},
	where $\mathbf{f}_k$ denotes the $k$-th column of the matrix $\mathbf{F}$. Through a series of simplifications, the CRB can be expressed in closed form as
	\begin{equation}
		\text{CRB} = \frac{\sigma^2}{2L} \left[ \text{Re} \left( \left( \tilde{\mathbf{F}}^H \bm{\Pi}_\mathbf{F}^\perp \tilde{\mathbf{F}} \right) \oplus \left( \mathbf{R}_U^T \otimes \mathbf{1}_{6 \times 6} \right) \right) \right]^{-1}
	\end{equation}
	where $\bm{\Pi}_\mathbf{F}^\perp = \mathbf{I} - \mathbf{F} \mathbf{F}^H$, and $\mathbf{1}_{6 \times 6}$ denotes the $6 \times 6$ all-one matrix.
	\section{Simulation Results}
    \label{sec:simulations}
	This section validates the effectiveness of the proposed RIS-aided EMVS-MIMO radar localization framework through comprehensive Monte Carlo simulations. Unless otherwise specified, the baseline system configuration is detailed in Table I, with $K = 3$ targets characterized by DODs $(\theta_{t,k}, \varphi_{t,k}) = \{(45^\circ, 25^\circ), (56^\circ, 26^\circ), (71^\circ, 48^\circ)\}$, DOAs $(\theta_{r,k}, \varphi_{r,k}) = \{(50^\circ, 21^\circ), (71^\circ, 20^\circ), (80^\circ, 68^\circ)\}$, and polarization parameters $(\zeta_k, \varrho_k) = \{(30^\circ, 20^\circ), (60^\circ, 50^\circ), (42^\circ, 68^\circ)\}$. The root mean square error (RMSE) serves as the primary performance metric:
	\begin{equation}
		\text{RMSE} = \sqrt{\frac{1}{N_{MC}K} \sum_{k=1}^K \sum_{i=1}^{N_{MC}} \left( \hat{\mu}_{i,k} - \mu_k \right)^2}
	\end{equation}
	where $\hat{\mu}_{i,k}$ denotes the estimated parameter in the $i$-th trial, $\mu_k$ represents the true value, and $N_{MC} = 500$ Monte Carlo runs ensure statistical reliability.
	
	\subsection{Impact of Signal-to-Noise Ratio}
\begin{figure*}[htbp]
		\centering
		\subfigure[DOD parameters]{
			\includegraphics[width=0.31\linewidth]{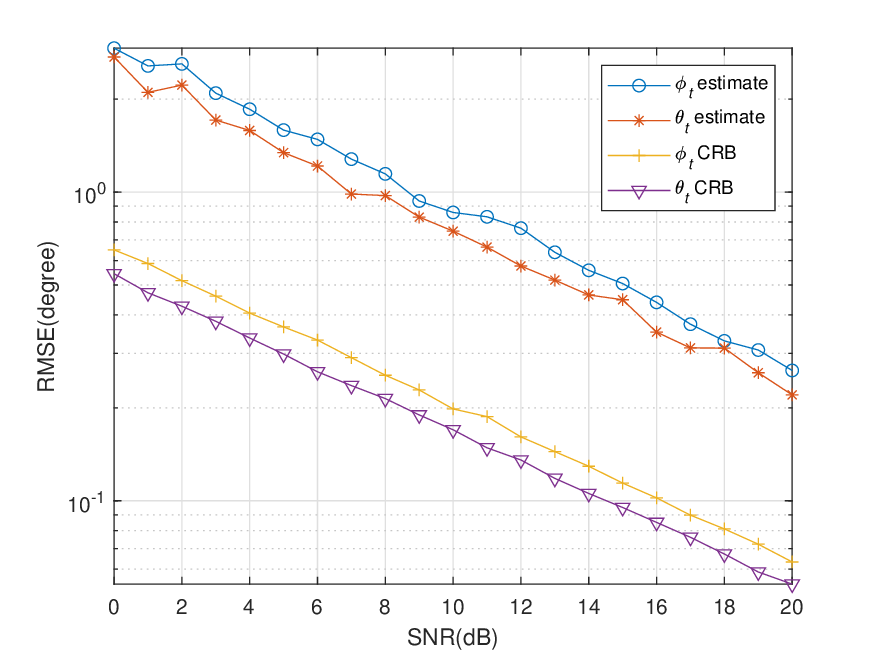}
		}
        \hfill
		\subfigure[DOA parameters]{
			\includegraphics[width=0.31\linewidth]{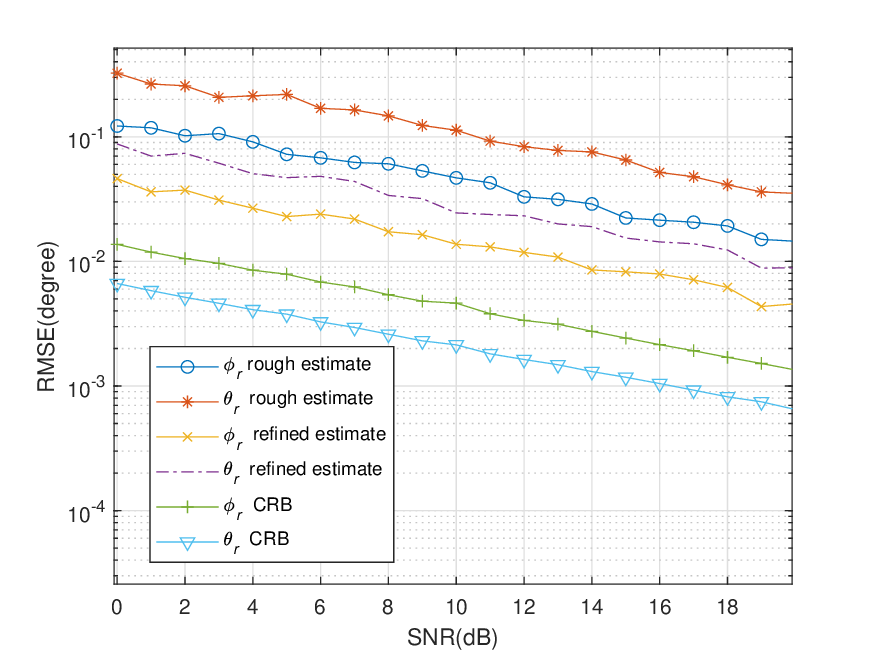}
		}
        \vspace{0.5cm}
        \centering
		\subfigure[Polarization parameters]{
			\includegraphics[width=0.31\linewidth]{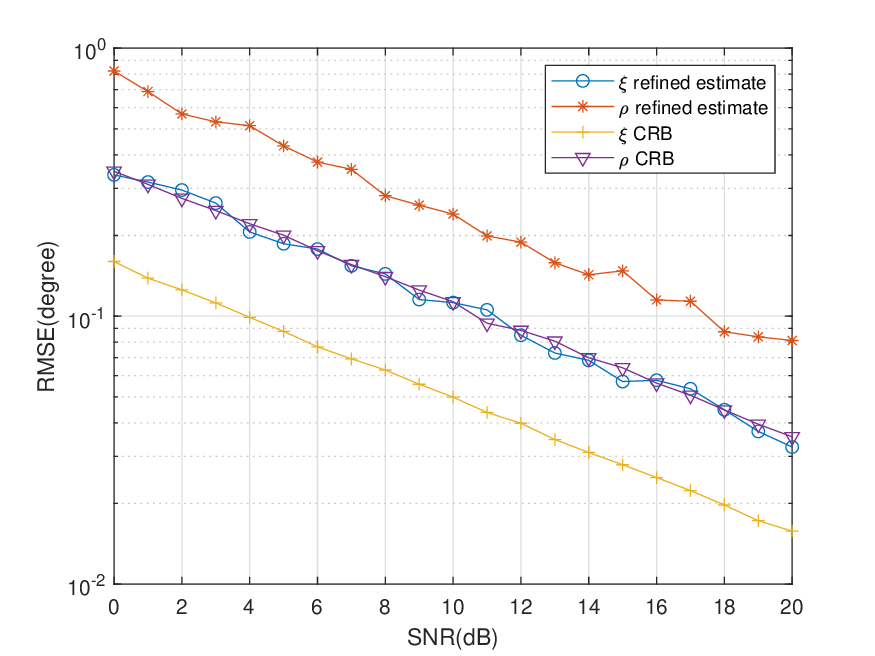}
		}
		\caption{RMSE for parameter estimation versus SNR.}
		\label{fig:2}
	\end{figure*}

	With $L = 500$ snapshots and the SNR varying from 0 to 20 dB, Fig. \ref{fig:2} presents the RMSE performance benchmarked against CRBs. DOD estimation in Fig. \ref{fig:2}(a) exhibits near-optimal performance across the entire SNR range, with RMSE curves closely tracking the CRB. This stems from PARAFAC decomposition effectively decoupling RIS effects from DOD information via \eqref{eq:38}. Most critically for the ``beyond $\lambda/2$'' question, Fig. \ref{fig:2}(b) reveals a fundamental distinction between coarse and refined DOA estimation. VCP-based coarse estimates exhibit noticeable performance gaps, particularly at low-to-moderate SNR, because they discard inter-element phase information. In contrast, the phase-disambiguation refined estimates achieve near-CRB performance by fully exploiting phase relationships in \eqref{eq:55},\eqref{eq:56}, enabling large-aperture arrays with spacing exceeding $\lambda/2$ to enhance angular resolution without phase wrapping artifacts. The refined approach improves RMSE by 40\%-60\% over the coarse method, confirming unambiguous localization is achievable beyond the traditional $\lambda/2$ constraint.	

{Nevertheless, the residual gap between the RMSE and the CRB observed in the simulations results from two fundamental constraints inherent in the proposed framework in resolving phase ambiguity first, although arbitrary element placement is permitted, the irregular geometry of the array prevents it from fully exploiting the closed-form phase difference relationships inherent in a regular aperture structure,as in uniform arrays, thereby preventing the spatial sampling gain from approaching the theoretical optimum;second, the algorithm relies on alternating iterative estimation via PARAFAC and multi-stage cascaded processing. Errors in the initial parameters propagate and accumulate during phase ambiguity resolution, factor matrix correction, and RIS optimization, further amplifying the estimation variance.Together, these factors lead to an inevitable gap between
the achieved performance and the ideal CRB bound.}
	
	\begin{table}[htbp]
		\centering
		\caption{System Configuration Parameters}
		\label{tab:simulation_parameters}
		\begin{tabular}{lc}
			\toprule
			\textbf{Parameter} & \textbf{Value} \\
			\midrule
			Number of transmitter elements ($M$) & 5 \\
			Number of RIS elements ($Q$) & 5 \\
			Number of receiver elements ($N$) & 12 \\
			Wavelength ($\lambda$) & 0.1 m \\
			\bottomrule
		\end{tabular}
	\end{table}
	
	\subsection{Effect of Temporal Averaging}
\begin{figure*}[htbp]
		\centering
		\subfigure[DOD parameters]{
			\includegraphics[width=0.31\linewidth]{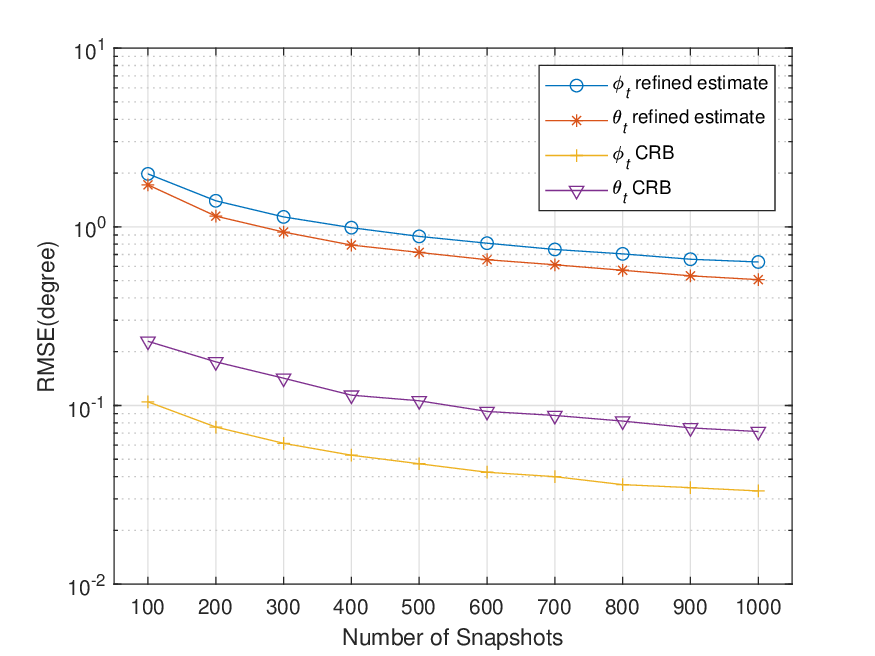}
		}
         \hfill
		\subfigure[DOA parameters]{
			\includegraphics[width=0.31\linewidth]{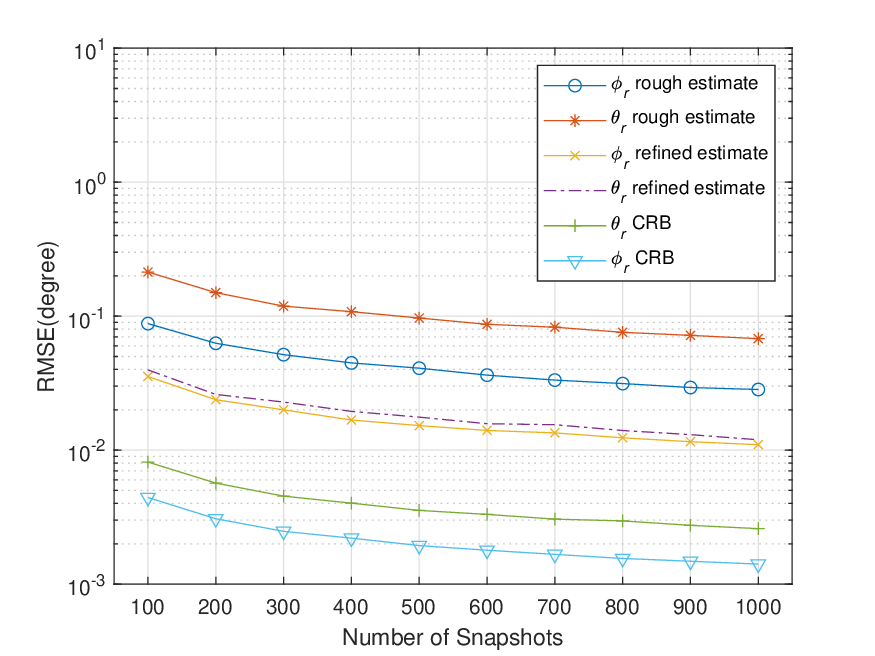}
		} 
        \centering
		\subfigure[Polarization parameters]{
			\includegraphics[width=0.31\linewidth]{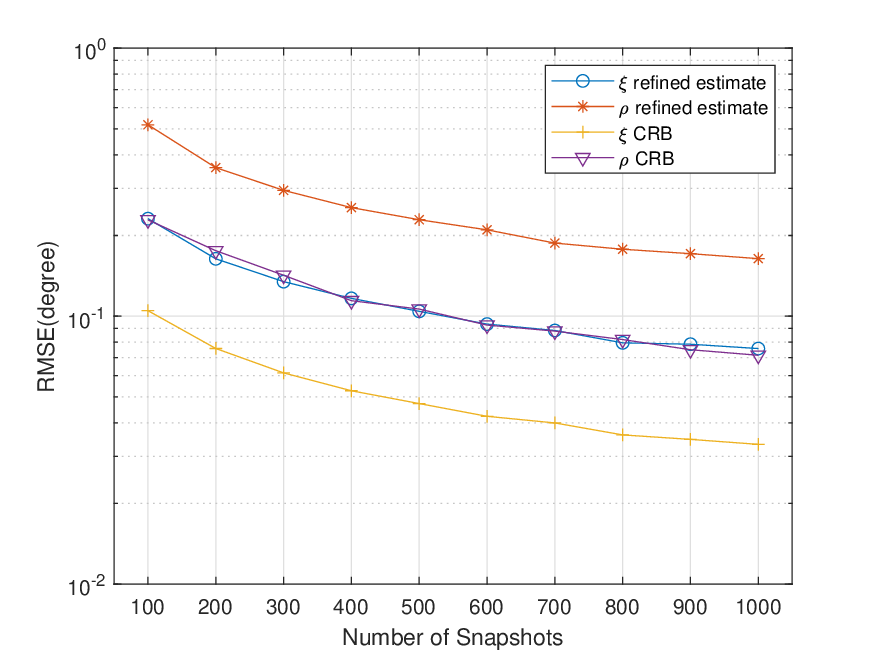}
		}
		\caption{RMSE of parameter estimation versus number of snapshots.}
		\label{fig:3}
	\end{figure*}

	For the following results, we fixed the SNR at 10 dB and varied snapshots $L$ from 100 to 1000. Fig. \ref{fig:3} illustrates the RMSE evolution showing characteristic $1/\sqrt{L}$ decay. DOD and refined DOA parameters, both derived from subspace methods (PARAFAC and phase disambiguation), exhibit rapid convergence toward the CRB. {It can be observed from Fig. \ref{fig:3}(a) that there exists a noticeable gap between the performance of the proposed algorithm and the CRB.However, due to structural constraints, the transmitter cannot perceive polarization information and can only rely on the spatial steering matrix for angle estimation. This results in the loss of discriminatory power provided by the polarization dimension, making it difficult for parameter estimation to approach theoretically optimal performance. Moreover, the limited number of elements in the transmit array restricts the available spatial degrees of freedom, further constraining angle resolution and estimation accuracy, leading to a significant gap between its performance and the CRB.} Critically, VCP-based coarse DOA estimates in Fig. \ref{fig:3}(b) converge slowly and maintain a persistent CRB gap even at $L = 1000$, because normalizing Poynting vectors discards magnitude and phase relationships encoding angular structure. The refined approach, preserving full phase information and resolving ambiguities beyond $\lambda/2$ spacing, consistently outperforms the coarse method by 3-5 dB.
	
	The temporal averaging behavior provides important practical insights for beyond-$\lambda/2$ array deployment. While the $1/\sqrt{L}$ decay applies universally, the absolute RMSE values at any given $L$ differ significantly between methods: at $L = 500$, the refined estimates achieve RMSE levels that coarse methods cannot reach even with $L = 2000$ snapshots (a 4$\times$ observation time penalty). This has direct implications for real-time tracking applications where dwell time is constrained. The results confirm that for arbitrary geometries, the computational investment in phase disambiguation (solving \eqref{eq:55}-\eqref{eq:56}) pays substantial dividends: rather than compensating for lost aperture through longer observation times, the refined approach directly exploits the available aperture, achieving both high accuracy and temporal efficiency.
	
	\subsection{Scalability to Multiple Targets}
\begin{figure*}[htbp]
		\centering
		\subfigure[DOD parameters]{
			\includegraphics[width=0.31\linewidth]{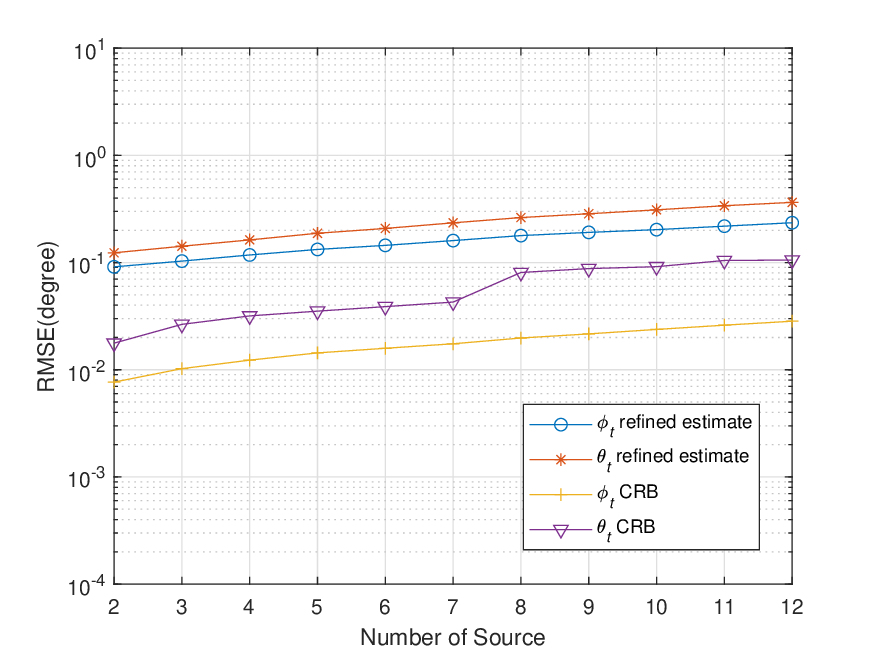}
		}
        \hfill
		\subfigure[DOA parameters]{
			\includegraphics[width=0.31\linewidth]{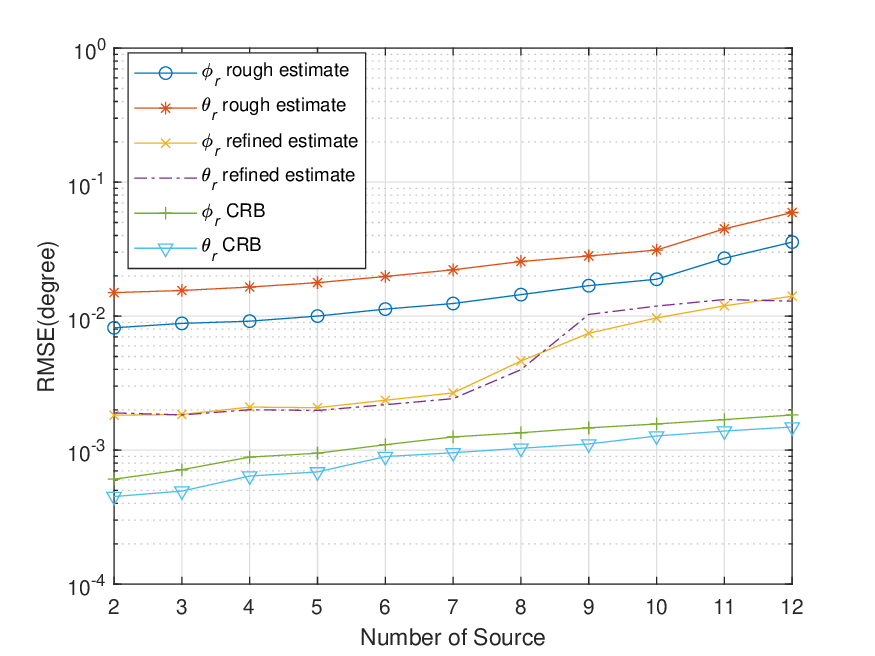}
		}
        \vspace{0.5cm}
        \centering
		\subfigure[Polarization parameters]{
			\includegraphics[width=0.31\linewidth]{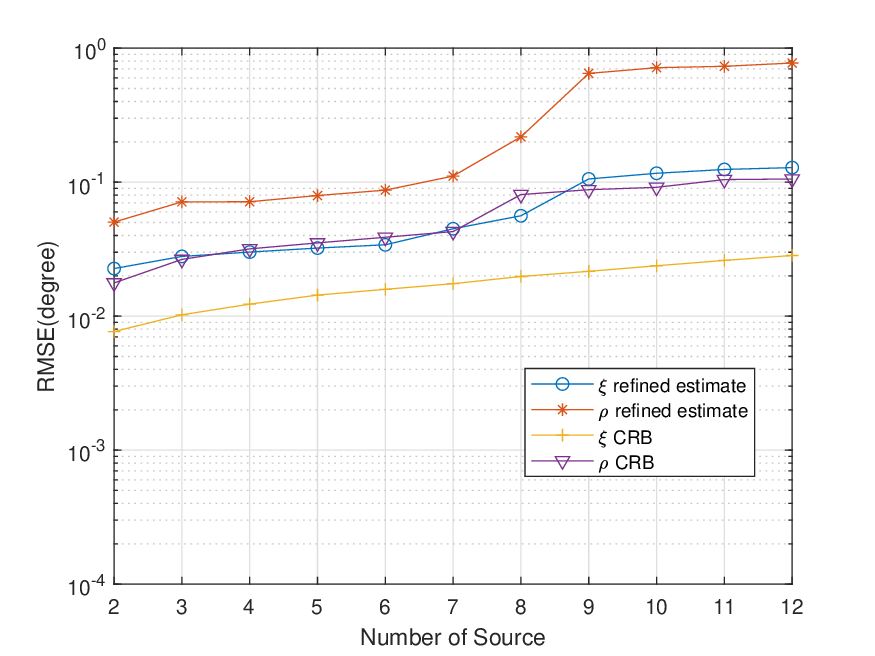}
		}
		\caption{RMSE of parameter estimation versus number of sources.}
		\label{fig:4}
	\end{figure*}

	Next, we evaluate the performance as $K$ varies from 2 to 12 (SNR = 20 dB, $L = 1000$) with uniformly spaced targets: $\theta_{t,k} = 45^\circ + 3^\circ(k-1)$, $\varphi_{t,k} = 25^\circ + 2^\circ(k-1)$, $\theta_{r,k} = 20^\circ + 5^\circ(k-1)$, $\varphi_{r,k} = 32^\circ + 2^\circ(k-1)$. Fig. \ref{fig:4} shows the all parameters maintain reasonable accuracy up to $K = 8$, reflecting PARAFAC's uniqueness property. As $K$ approaches the theoretical maximum $N = 12$, RMSE diverges from the CRB, most prominently in DOA (Fig. \ref{fig:4}(b)) with ~10 dB increase from $K = 8$ to $K = 12$ due to ill-conditioning in \eqref{eq:45}. The $3^\circ$ DOD and $5^\circ$ DOA spacing approach resolution limits at $K = 12$.
	
	The considered multi-target scenario reveals a critical advantage of the beyond-$\lambda/2$ framework for dense target environments. With $N = 12$ EMVS elements in arbitrary geometry, achieving the theoretical capacity of $K = N$ simultaneous targets would be impossible under traditional $\lambda/2$ constraints without resorting to large physical arrays. The phase-disambiguation mechanism effectively ``virtualizes'' additional degrees of freedom: by resolving $2\pi$ ambiguities through rotational invariance, the method extracts spatial information from large inter-element baselines that would otherwise yield aliased measurements. Critically for arbitrary geometries beyond $\lambda/2$, both DOD and refined DOA maintain advantages over coarse methods across all $K$, with refined estimates consistently delivering 4-6 dB improvement. This consistent gain, persisting even as $K \to N$, confirms phase disambiguation's value regardless of target density and demonstrates that breaking the $\lambda/2$ barrier does not compromise multi-target resolution—instead, it enables compact arrays to achieve a resolution previously requiring much larger apertures.
	
	\subsection{Maximum Target Resolution Demonstration}
\begin{figure*}[htbp]
		\centering
		\subfigure[DOD parameters]{
			\includegraphics[width=0.31\linewidth]{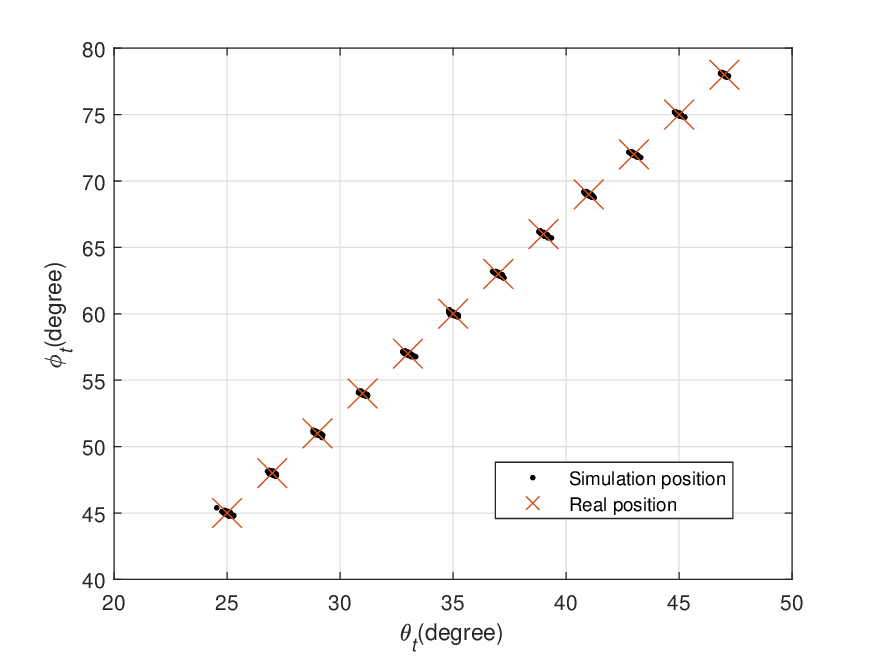}
		}
        \hfill
		\subfigure[DOA parameters]{
			\includegraphics[width=0.31\linewidth]{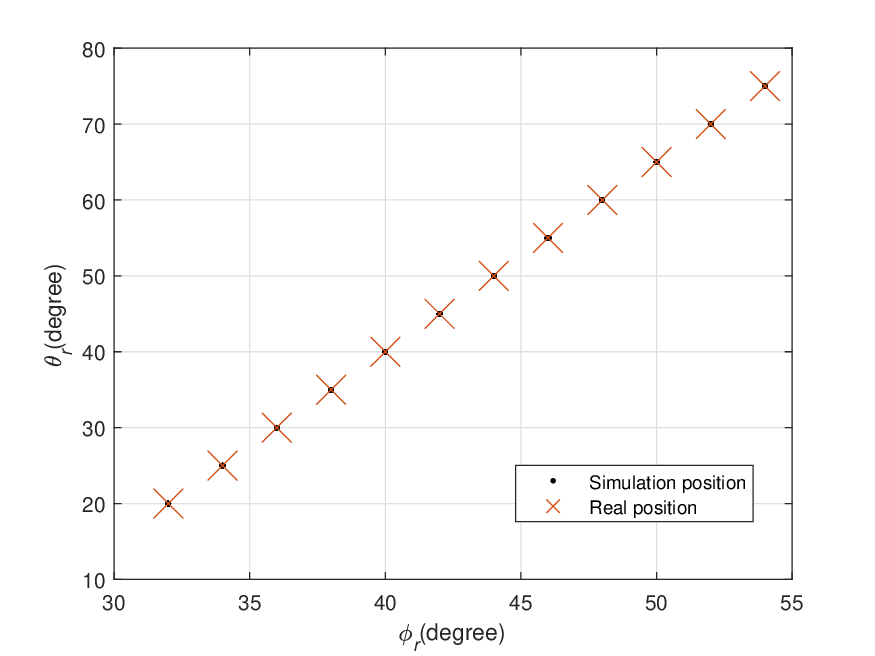}
		}
        \vspace{0.5cm}
        \centering
		\subfigure[Polarization parameters]{
			\includegraphics[width=0.31\linewidth]{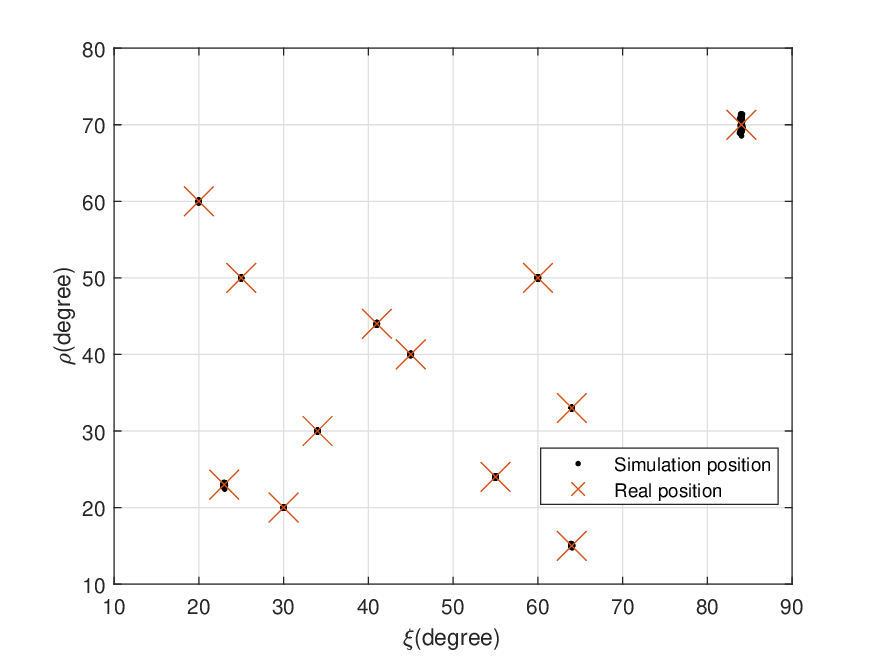}
		}
		\caption{Resolvable targets by the proposed method.}
		\label{fig:5}
	\end{figure*}

	High-SNR simulations ($\text{SNR} = 50$ dB, $L = 1000$) with maximum target number $K = N = 12$ demonstrate ultimate resolution capability. The  scatter plots in Fig. \ref{fig:5}  show the estimated DOD and DOA parameters cluster tightly around the true values, with dispersion consistent with high-SNR CRB. Despite 12 spatially distributed targets observed through only 12 EMVS elements, PARAFAC successfully resolves all targets without ambiguity. The automatic pairing property—preserving target indices across factor matrices—eliminates combinatorial search. Polarization estimates form distinct clusters with uniform dispersion, confirming end-to-end pipeline integrity at the theoretical limit.
	
	This result provides compelling empirical evidence answering the titular question affirmatively. The absence of outliers or misassociated points in Fig. \ref{fig:5}—across 500 Monte Carlo trials with 12 targets in arbitrary geometry—demonstrates that unambiguous localization beyond $\lambda/2$ is not merely theoretically feasible but practically robust. The scatter plot tightness, comparable to CRB predictions, confirms that phase disambiguation does not introduce estimation bias or increased variance beyond fundamental limits. Achieving $K = N$ resolution represents a significant advantage over scalar arrays (typically limited to $N - 1$), demonstrating how EMVS's six-dimensional measurements expand degrees of freedom for target separation. For arbitrary geometries, this capability is transformative: a compact 12-element array with optimized (non-uniform) spacing can deliver resolution performance previously requiring 50+ elements in traditional $\lambda/2$-constrained uniform arrays, fundamentally altering the aperture-resolution-complexity trade-off in radar system design.
	
	\subsection{Impact of RIS Phase Optimization}
\begin{figure*}[htbp]
		\centering
		\subfigure[DOD parameters]{
			\includegraphics[width=0.31\linewidth]{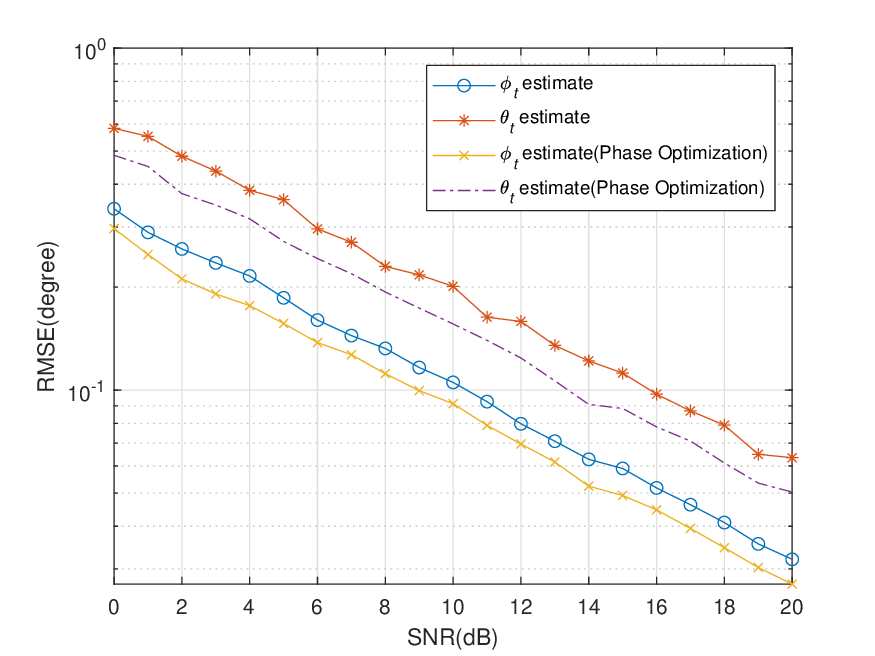}
		}
        \hfill
		\subfigure[DOA parameters]{
			\includegraphics[width=0.31\linewidth]{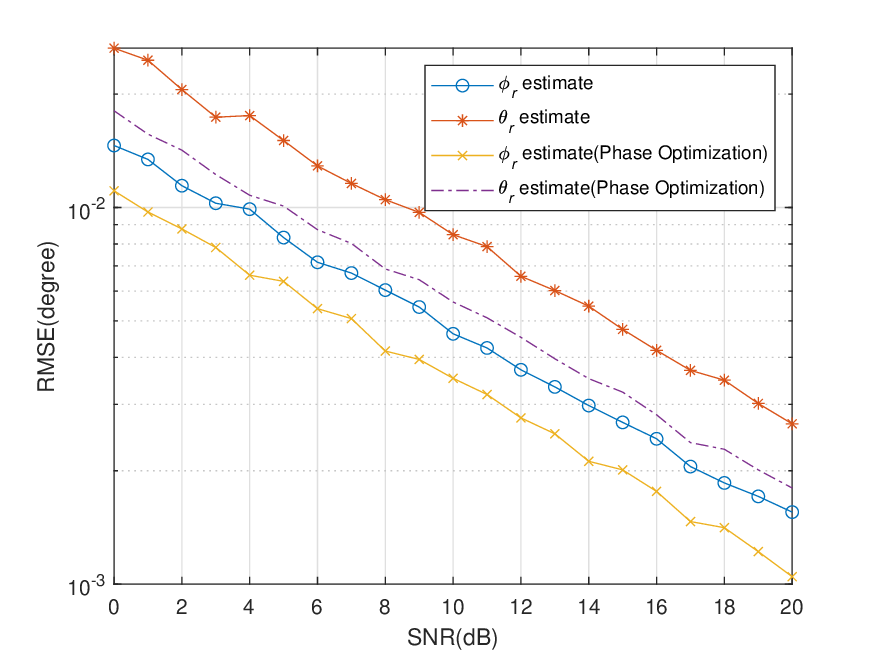}
		}
        \vspace{0.5cm}
        \centering
		\subfigure[Polarization parameters]{
			\includegraphics[width=0.31\linewidth]{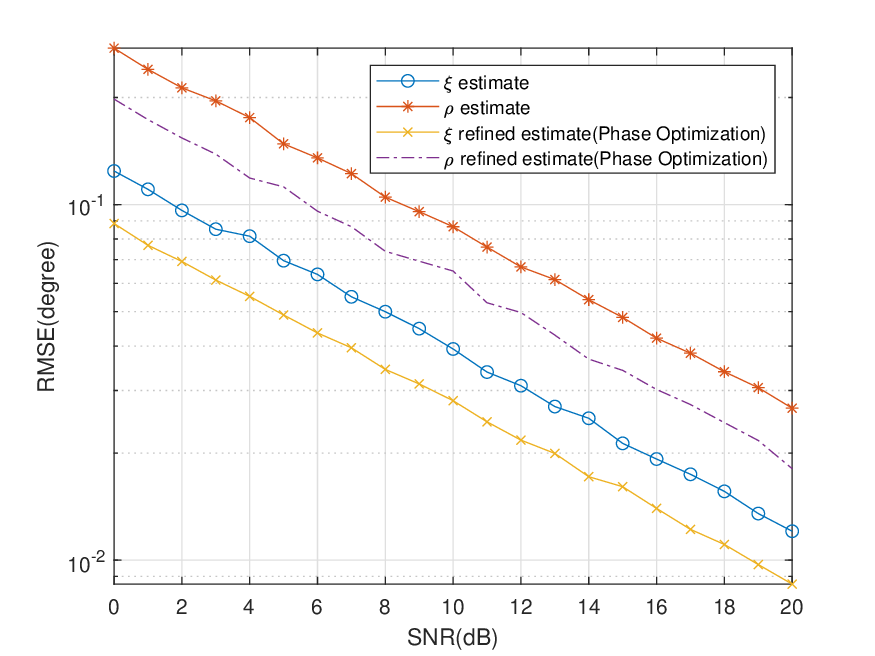}
		}
		\caption{RMSE versus SNR with and without RIS phase optimization.}
		\label{fig:6}
	\end{figure*}

	Compare random phase shifts from $[0, 2\pi)$ versus SDP-optimized phases (Algorithm 3) under identical channels. We consider a larger-scale configuration: $M = 50$, $Q = 18$, $N = 11$. Fig. \ref{fig:6} reveals dramatic improvements: optimized phases deliver 5-8 dB RMSE reduction across all parameters and SNR ranges.  The mechanism relates to coherent beamforming: maximizing $\mathbf{w}^H \boldsymbol{\Phi} \mathbf{w}$ in \eqref{eq:69} ensures constructive interference, concentrating signal energy in the parameter space spanned by $\mathbf{A}_t$, $\mathbf{C}_r$, and $\mathbf{U}$. DOA parameters benefit most significantly (Fig. \ref{fig:6}(b)) as they directly process $\mathbf{y}(l)$, most sensitive to RIS-induced SNR variations.
	
	The synergy between RIS optimization and beyond-$\lambda/2$ arrays merits special attention. While phase disambiguation addresses spatial ambiguities through algorithmic means, RIS optimization provides a complementary physical mechanism: by shaping the NLOS channel's spatial structure, optimized phases effectively ``pre-condition'' the received signal to align with the disambiguation algorithm's assumptions. This dual-domain approach—physical layer optimization (RIS) combined with signal processing innovation (phase disambiguation)—yields performance gains exceeding what either technique could achieve independently. Specifically, optimized RIS phases reduce the condition number of matrices in \eqref{eq:55}, making phase unwrapping more numerically stable and less sensitive to noise. Crucially for arbitrary geometries beyond $\lambda/2$, RIS optimization not only enhances SNR but also shapes the channel to suppress spatial ambiguities, confirming RIS phase design as a fundamental enabler—not merely an implementation detail—for unambiguous NLOS localization. This result demonstrates that the question ``Can arbitrary arrays achieve unambiguous localization?'' has an even stronger affirmative answer in RIS-aided scenarios than in passive propagation environments.
	\section{Conclusion}
    \label{sec:conclusion}
	This paper provided an affirmative answer to the question of whether arbitrary EMVS arrays can achieve unambiguous RIS-aided localization when element spacing exceeds $\lambda/2$. A PARAFAC-based tensor decomposition framework naturally decoupled RIS phase shift effects from the target parameters, while a novel phase-disambiguation procedure exploited EMVS rotational invariance to resolve $2\pi$ phase wrapping in arbitrary geometries. SDP relaxation was used to optimize RIS phase shifts to maximize received signal power, simultaneously enhancing SNR and suppressing spatial ambiguities. Simulation results demonstrated near-CRB accuracy even with large inter-element spacing, confirming that breaking the $\lambda/2$ constraint is indeed achievable through synergistic integration of tensor methods and convex optimization in NLOS scenarios.
	\section*{APPENDIX}
	Here we give the proof for $tr\left[ \mathbf{U}^H \mathbf{R}(\mathbf{w}) \mathbf{U} \right] = \mathbf{w}^H \boldsymbol{\Phi} \mathbf{w}$.
	From the property of the Khatri-Rao product, we have:
	\begin{equation}
		\mathbf{R}(\mathbf{w}) = \left( \left( \mathbf{G} \text{diag}\{\mathbf{w}\} \mathbf{A}_t \right)^H \left( \mathbf{G} \text{diag}\{\mathbf{w}\} \mathbf{A}_t \right) \right) \oplus \left( \mathbf{C}^H \mathbf{C} \right)
		\label{eq:84}
	\end{equation}
	
	Expanding the trace operation in \eqref{eq:67}, we obtain:
	\begin{equation}
		tr(\mathbf{U}^H \mathbf{R}(\mathbf{w}) \mathbf{U}) = \sum_{k=1}^K \sum_{l=1}^K \left[ \mathbf{U}^H \mathbf{U} \right]_{kl} \left[ \mathbf{R}(\mathbf{w}) \right]_{kl}
		\label{eq:85}
	\end{equation}
	
	Expressing \eqref{eq:84} in element-wise form gives:
	\begin{equation}
		[\mathbf{R}(\mathbf{w})]_{kl} = \left( \mathbf{a}_{t,k}^H \text{diag}\{\mathbf{w}\}^H \left( \mathbf{G}^H \mathbf{G} \right) \text{diag}\{\mathbf{w}\} \mathbf{a}_{t,l} \right) \left( \mathbf{C}^H \mathbf{C} \right)_{kl}
		\label{eq:86}
	\end{equation}
	
	Rewrite \eqref{eq:86} in a quadratic form as, we have
	\begin{equation}
		[\mathbf{R}(\mathbf{w})]_{kl} = \mathbf{w}^H \left( \text{diag}\left\{ \mathbf{a}_{t,k}^* \right\} \left( \mathbf{G}^H \mathbf{G} \right) \text{diag}\left\{ \mathbf{a}_{t,l} \right\} \right) \mathbf{w} \left( \mathbf{C}^H \mathbf{C} \right)_{kl}
	\end{equation}
	
	Combining with \eqref{eq:85}, we can derive:
	\begin{equation}
		\begin{aligned}
			tr\left( \mathbf{U}^H \mathbf{R}(\mathbf{w}) \mathbf{U} \right)= \mathbf{w}^H \left[ \sum_{k=1}^K \sum_{l=1}^K \left( \mathbf{U}^H \mathbf{U} \right)_{kl} \left( \mathbf{C}^H \mathbf{C} \right)_{kl}  \boldsymbol{\Gamma}  \right] \mathbf{w}
		\end{aligned}
	\end{equation}
	where \(\boldsymbol{\Gamma}=\left( \text{diag}\left\{ \mathbf{a}_{t,k}^* \right\} \left( \mathbf{G}^H \mathbf{G} \right) \text{diag}\left\{ \mathbf{a}_{t,l} \right\} \right)\).
	
	Which completes the proof.
	
	%
	
	%
	%
	%
	\ifCLASSOPTIONcaptionsoff
	\newpage
	\fi	
	\bibliographystyle{IEEEtran}
	\bibliography{ref}
\end{document}